\documentclass{aa}
\usepackage{txfonts}
\usepackage{amssymb}
\usepackage{graphicx}
\usepackage{hyperref}
\def\asec{\ifmmode ^{\prime\prime}\else$^{\prime\prime}$\fi}

\def\it{\sl}
\def\degs{\ifmmode ^{\circ}\else$^{\circ}$\fi}
\def\amin{\ifmmode ^{\prime}\else$^{\prime}$\fi}
\def\asec{\ifmmode ^{\prime\prime}\else$^{\prime\prime}$\fi}
\def\fm{\hbox{$.\!\!^{\rm m}$}}            % Fractions of magnitudes
\def\farcs{\hbox{$.\!\!^{\prime\prime}$}}  % Fractions of arcseconds
\def\psr{PSR~B1133+16}
\def\degs{\ifmmode ^{\circ}\else$^{\circ}$\fi}
\def\amin{\ifmmode ^{\prime}\else$^{\prime}$\fi}
\def\farcm{\hbox{$.\mkern-4mu^\prime$}}
\def\eqalign#1{\null\,\vcenter{\openup1\jot \m@th
   \ialign{\strut\hfil$\displaystyle{##}$&$\displaystyle{{}##}$\hfil
   \crcr#1\crcr}}\,}
\begin{document}
\authorrunning{S. Zharikov, et al.}
\titlerunning{ Optical observations of the \psr}
\title{
Possible optical detection of a fast, 
%moving 
nearby radio pulsar PSR~B1133+16\thanks{Based 
on observations made with ESO telescope at the Paranal Observatory under
Program 71.D-0499(A) and with archival ESO VLT data, obtained from
the ESO/ST-ECF Science Archive Facility.}}
\author{S.V.~Zharikov\inst{1}
\and Yu.A.~Shibanov\inst{2} 
\and R.E.~Mennickent\inst{3}
\and V.N.~Komarova\inst{4,5}
}
\offprints{S. Zharikov, \\ \email{ zhar@astrosen.unam.mx}}
\institute{
Observatorio Astron\'{o}mico Nacional SPM, Instituto de Astronom\'{i}a, Universidad Nacional Aut\'{o}nomia de M\'{e}xico, Ensenada, BC, M\'{e}xico, 
zhar@astrosen.unam.mx
\and
Ioffe Physical Technical Institute, Politekhnicheskaya 26,
St. Petersburg, 194021, Russia \\
shib@astro.ioffe.ru
\and
Departamento de Fisica, Universidad de Concepcion, Casilla 160-C, Concepcion, Chile \\
rmennick@stars.cfm.udec.cl 
\and 
Special Astrophysical Observatory Russian Academy of Science, Nizhnii Arkhyz, 
Russia, 369167 \\
vkom@sao.ru
\and Isaac Newton Institute of Chile, SAO Branch, Russia 
}

\date{Received ---  , accepted --- } 
%%%%%%%%%%%%%%%%%%%%%%%%%%%%%%%%%%%%%%%%%%%%%%%%%%%%%%%%%%%
%%%%%%%%%%%%%%%%%%%%  ABSTRACT %%%%%%%%%%%%%%%%%%%%%%%%%%%%
%%%%%%%%%%%%%%%%%%%%%%%%%%%%%%%%%%%%%%%%%%%%%%%%%%%%%%%%%%%
\abstract{}
{We 
performed  deep optical  
observations of the 
field of an old, fast-moving radio pulsar PSR B1133+16 in  an attempt  to detect its optical counterpart and a bow shock
nebula.  }{The observations were carried out using the direct imaging mode of
 FORS1 at the ESO VLT/UT1 telescope in the $B$, $R$, and  $H_\alpha$ bands. 
We also used  
archival  images of the same 
field obtained with the VLT in the $B$ band and with the Chandra/ACIS in X-rays.   
}
{ In the $B$ band we detected  a faint
($B$=28\fm1$\pm$0\fm3)  source that may be the  optical counterpart 
of PSR B1133+16, 
as it is positionally consistent with the radio pulsar and
with the  X-ray  counterpart candidate published earlier.
Its upper limit in the $R$ band 
implies
a color
index $B$-$R$$\la$0\fm5,  
  which is compatible with the index values  for    
  most pulsars 
 identified in the optical range. 
The derived optical luminosity and its ratio to the X-ray luminosity
of the candidate are consistent  with  expected values
derived  from a sample of pulsars detected in both spectral domains. 
No  Balmer bow shock  was detected, implying a low density 
of  ambient matter around 
the pulsar.  However,   in the X-ray and  
$H_\alpha$ images we found  the signature of a  trail  extending   $\sim 4\asec-5\asec$ behind the pulsar and
coinciding with the direction of its proper motion. If confirmed by deeper studies, 
this is the first time 
such a trail has been seen in the optical and X-ray wavelengths.
}      
{Further  
observations 
at later epochs are necessary  to confirm the identification of the pulsar
by the candidate's proper motion measurements.
}

\keywords{pulsars:   general    --   pulsars,   individual:    PSR B1133+16  --
stars: neutron} 
\maketitle
%
%%%%%%%%%%%%%%%%%%%%%%%%%%%%%%%%%%%%%%%%%%%%%%%%%%%%%%%%%%%
%%%%%%%%%%%%%%%%%%%%  SECTION  %%%%%%%%%%%%%%%%%%%%%%%%%%%%
%%%%%%%%%%%%%%%%%%%%%%%%%%%%%%%%%%%%%%%%%%%%%%%%%%%%%%%%%%%
%%%%%%%%%%%%%%%%%%%%%%%%%%%%%%%%%%%%%%%%%%%
\section{Introduction}
\label{sec1}
%%%%%%%%%%%%%%%%%%%%%%%%%%%%%%%%%%%%%%%%%%%%
Until now optical emission 
%was
has only been   
detected  %only 
from  $\le1\%$ of
$\ga$1500 known radio pulsars  (e.g.,  Mignani et al. \cite{Mignani2005}).  
Even such a small number of 
optical 
identifications %detections shows
implies 
that  rotation-powered neutron stars (NSs) can
be  active in the  optical,  % lifetime    
as  
%long 
well
as  in the radio range. %pulsars.       
This follows from the fact that a small group of the optical pulsars  
%contains not only of the young, $\sim$1~kyr, and energetic objects
contains not only young, $\sim$1~kyr, and energetic objects
%like Crab  (Percival et al.~1993),  but also  of
like the Crab pulsar (Percival et al.\cite{Percival}),  but also
much older  pulsars  such
%{ pulsars as old as}
as  $\sim$3.1~Myr  \object{PSR
B1929+10} (Pavlov et al. \cite{Pavlov}; Mignani et al.
\cite{Mignani2002}),  $\sim$17.4~Myr \object{PSR B0950+08}
(Pavlov et al. \cite{Pavlov}; Zharikov et al.
\cite{zhar2002, zhar2004}),  %as well as  %and 
{ and even the}
%a 
very old $\sim$1~Gyr recycled millisecond pulsar \object{PSR J0437-4715} 
(Kargaltsev et al. \cite{kar2004}).
The  power-law non-thermal 
spectral component is dominant in the  pulsar's optical emission, 
presumably originating in the magnetospheres of the NSs.

\begin{table*}[t]
\caption{Parameters of several nearby  
old pulsars.} 
\begin{tabular}{cccccccccc} \hline \hline
Pulsar       & age & distance$^*$ & parallax$^*$ & $\mu_\alpha^*$ & $\mu_\delta^*$ &{\it v}$_{\perp}^*$ 
& log($\dot{E}$) & optical band & magnitude \\ 
     & log ($\tau$ ) & pc       & mas     &mas yr$^{-1}$&mas yr$^{-1}$& km s$^{-1}$ & ergs s$^{-1}$ & &        \\ \hline \hline
PSR B1929+10 & 6.49    & 331(10)  & 3.02(9) &  94.8   &43 & 163(5)   & 33.59    & F342W & 26.8$^a$   \\
{ PSR B1133+16} & 6.70    & 350(20)  & 2.80(16)& -73.95   &368.05&
631(38)  & 31.94    & - &  28.1$^d$         \\ 
PSR B0950+08 & 7.24    & 262(5)   & 3.82(7) & -2.09   &29.5& 36.6(7)  & 32.74    &  B &  27.07$^b$     \\ 
PSR J0108-1431& 8.23   & 60-130   & -      & $<26$ & $<78$ & $<50_{130}$& 30.75  & B &$ >28.6^c  $           \\  \hline 
\end{tabular}
\begin{tabular}{l}
$^*$ Brisken et al.~\cite{Brisken}; 
$^a$ Pavlov et al.~\cite{Pavlov};
$^b$ Zharikov et al.~\cite{zhar2002, zhar2004};  
$^c$ Mignani et al.~\cite{Mignani03};
$^d$ this paper   
\end{tabular}
\label{T.1}
\end{table*} 
%%%%%%%%%%%%%%%%%%%%%%%%%%%%%%
\begin{table*}[tbh]
\caption{Log of the VLT/FORS1 observations of the
\object{\psr} field 
during  the  2003-2004 periods}
\begin{tabular}{cclcccc|cclcccc}
\hline\hline
 Date       &       UT  &Band, exp.  & Exp.&
 secz& Seeing & Sky & Date      &   
    UT  &Band, exp. & Exp.& secz& Seeing & Sky\\ 
  UT        &           & number    &  $sec.$   &
     &$arcsec.$ &$\frac{ counts}{sec\ pix}$ & UT        &     
          &number     &  $sec.$   &   &$arcsec.$ & $\frac{counts}{sec\ pix}$\\
\hline 
03 Apr 2003 & 02:59     & B 1  & 1020     & 1.33   &  0.95 & 3.04&  -''-       & 08:03     & B 12             & 1020     & 1.57   &  0.97 & 2.53\\
 -''-       & 03:17     & B 2  & 1020     & 1.31   &  0.92 &3.02& 25 Dec 2003 & 07:56     & R  7            & 485      & 1.58    &  0.57 &17.38\\
 -''-       & 03:35     & B 3  & 1020     & 1.31   &  0.91 &2.98&  -''-       & 08:05     & R  8            & 485      & 1.54    &  0.55 &16.80\\
 -''-       & 03:55     & R 1  & 750      & 1.32   &  0.77 &27.24  &  -''-       & 08:15     & B 13             & 30	& 1.50     &  1.01 &-\\ 
 -''-       & 04:08     & R 2  & 750      & 1.33   &  0.71
 &25.78& 19 Jan 2004 & 07:24     & H$_\alpha$  1   & 1550	& 1.37     &  0.67 &0.65\\
 -''-       & 04:22     & R 3  & 750      & 1.34   &  0.71
 &24.75&  -''-       & 07:51     & H$_\alpha$  2   & 1550	& 1.33     &  0.64 &0.67\\ 
 -''-       & 04:35     & R 4  & 750      & 1.37   &  0.73
 &24.21& 25 Jan 2004 & 07:08     & H$_\alpha$  3   & 1550	& 1.35     &  0.83 &0.71\\
04 Apr 2003 & 03:33     & R 5  & 750      & 1.31   &  0.91
&17.16&  -''-       & 07:35     & H$_\alpha$  4   & 1550	& 1.32     &  0.89 &0.74\\
 -''-       & 03:46     & R 6  & 750      & 1.32   &  0.84
 &17.47& 26 Jan 2004 & 05.46     & H$_\alpha$  5   & 1550	& 1.61     &  0.93 &0.59\\
09 Apr 2003 & \underline{02:37} & B 4  & 1020     & 1.32   & 
0.87 &5.24&  -''-       & 06:13     & H$_\alpha$ 6    & 1550	& 1.48     &  0.83 &0.57\\ 
 -''-       & \underline{02:55} & B 5  & 1020     & 1.31   &  0.86 &4.14&  -''-       & 06:41     & $H_\alpha$ 7    & 1550     & 1.40	   &  0.99 &0.59\\ 
 -''-       & 03:12     & B 6  & 1020     & 1.31   &  0.95
 &3.08&   -"-       & 07:08     & H$_\alpha$  8   & 1550     & 1.35	   &  1.11 &0.57\\
 -''-       & 03:32     & B 7  & 1020     & 1.32   &  0.95
 &2.65& 27 Jan 2004 & 07:22     & H$_\alpha$  9   & 1550     & 1.32	   &  0.88 &0.75\\ 
 -''-       & 03:49     & B 8  & 1020     & 1.33   &  0.81 &2.62& 31 Jan 2004 & 07:03     & R   9           & 750      & 1.33	   &  0.56 &23.42\\
 -''-       & 04:08     & B 9  & 1020     & 1.36   &  0.79 &2.70&  -''-       & 07:17     & R   10           & 750      & 1.32	   &  0.66 &15.52\\
24 Dec 2003 & 07:28     & B 10  & 1020     & 1.79   &  0.94 &2.56&  -"-        & 07.30     & R  11            & 750      & 1.31	   &  0.89 &15.15\\
 -''-       & 07:46     & B 11  & 1020     & 1.67   &  0.88 &2.52&  -''-       & 07:43     & R  12            & 750      & 1.31	   &  0.83 &14.95\\ 
\hline
\end{tabular}
\label{t:log}
\end{table*}
%%%%%%%%%%%%%%%%%%%%%%%%%%%%%%%%%%%%%%%%%%%%%%%%%%%%%%%%%%  %It is commonly accepted that 
The non-thermal component, which also is frequently observed  in 
a wider spectral range including  X-rays,   
  %emission of pulsars 
is believed to be powered by the NS rotational energy 
loss $\dot{E}$, called a spin-down luminosity.    
The parameter $\eta = L/\dot{E}$, where $ L$ is the radiative luminosity, 
describes the efficiency of the transformation of the rotational energy 
into the emission.         
The most striking result of  the optical study  
of   ordinary pulsars 
is 
%the fact 
that     
%from
%magnetospheres 
%of both old  
the two old pulsars, \object{PSR B1929+10} 
and \object{PSR B0950+08}, with rather low $\dot{E}$, 
have  almost { the same optical efficiency %ies   
%vk as the young } and energetic Crab-like objects with much higher
as young } and energetic Crab-like objects with much higher
spin-down luminosities (Zharikov et al.~\cite{zhar2004}).       
At the same time, the efficiencies of middle-aged
pulsars are significantly lower.   
%Here $ L_{opt}$ and $\dot{E}$  are the pulsar optical
%and spindown luminosities, respectively. 
In addition, a strong  correlation 
between the  non-thermal optical and X-ray  luminosities
of pulsars  was found, indicating %on 
a general origin 
of the emission in both ranges.  
(Zharikov et al. \cite{zhar2004, zhar2006}).  
%vk Increasing the number of the optically identified  pulsars
% is  necessary to confirm these findings.
To confirm these findings it is necessary to
increase the number of optically identified  pulsars.
%  one has 
%to increase the number  of old pulsars  identified in the
%optical range.   

An old, $\sim$5~Myr, nearby pulsar \object{PSR B1133+16}
has almost the same parameters as { the two old objects  mentioned above} 
(see Table~\ref{T.1}). 
  It is located at a high galactic latitude, 
 $l=242\degr,\
 b=69\degr$,
 implying a low interstellar extinction $E(B-V)=0.04$
(Schlegel et al. \cite{Schlegel}).  The direct proper
motion and  annual parallax 
measurements  in the radio range  by Brisken et al.
 (\cite{Brisken})  
%vk provide
yield
a high transverse velocity of 631$\pm$30~km/s  
and a short distance to the pulsar %of
 350$\pm$20~pc.
 The pulsar is younger than  \object{PSR B0950+08}, 
 but its spin-down luminosity,  
$8.8 \times 10^{31}$ ergs s$^{-1}$, is about an order of magnitude
lower than those  of  \object{PSR B1929+20} and \object{PSR B0950+08}. 
Nevertheless, it is higher than that of %another
the nearby old PSR J0108-1431, 
whose optical counterpart has been unsuccessfully searched 
by Mignani et al.~(\cite{Mignani03}).    
The high transverse velocity  is promising for detecting      
an $H_{\alpha}$ bow shock nebula expected to be produced by  the supersonic 
motion  of the pulsar in the interstellar matter, 
as has been found around several %other 
rapidly moving pulsars 
and radio-silent NSs (e.g., Gaensler \& Slane \cite{GaeSlan}).
Recently,  the %pulsar 
field of PSR B1133+16 has been observed in X-rays  with 
the Chandra/ACIS %using about 18 ks exposure  
and a faint %to the
  X-ray counterpart candidate  was found 
 at a flux level of (8$\pm$2)$\times
10^{-15}~$erg~s$^{-1}$cm$^{-2}$ in 0.5~--~8 keV range (Kargaltsev et al.~\cite{Karga06}). 
Given 
this value
and using an empirical  relation  between the optical 
and X-ray luminosities of pulsars (Zharikov et al.~\cite{zhar2004},~\cite{zhar2006}; 
see also Zavlin \& Pavlov \cite{Zav2004}),  one can expect to detect 
the  pulsar in the optical range at  a sensitivity level of 
27--29 magnitudes.  

 In this paper 
we present   
the results of a 
deep optical imaging
of the \object{PSR B1133+16} field taken  with 
the ESO Very Large Telescope (VLT)  in  $B$, $R_c$, and $H_\alpha$
bands to search for the optical counterpart and  
the 
bow shock nebula of the pulsar. 
We  also used
archival VLT images  of the field obtained earlier  in
the $B$ band and the X-ray data  from the Chandra archive. 
The observations and  data reduction are described  in
Sect.~\ref{sec2}.   Astrometric and photometric referencing   
are given  in Sect.~\ref{sec3}. We present our results  in Sect.~\ref{sec4} and 
discuss them  in  Sect.~\ref{sec5}. 

%%%%%%%%%%%%%%%%%%%%%%%%%%%%%%%%%%%%%%%%%%%%%%%%%%%%%%%%%%%
%%%%%%%%%%%%%%%%%%%%  SECTION  %%%%%%%%%%%%%%%%%%%%%%%%%%%%
%%%%%%%%%%%%%%%%%%%%%%%%%%%%%%%%%%%%%%%%%%%%%%%%%%%%%%%%%%%
\section{Observations and data reduction}  
%Archive Data}
\label{sec2}

The observations  
were carried out with the FOcal Reducer and low-dispersion Spectrograph 
 (FORS1\footnote{For the instrument details see {www.eso.org/instruments/fors/}})
at  the UT1 (ANTU)  unit of the ESO/VLT  during several
service-mode observational runs 
from the beginning of
April 2003 
to the end of January 2004. 
 A standard resolution 
 mode was used with an image scale of  
 $\sim$0\farcs2/$\mathrm{pixel}$  and field of view
 (FOV) $\sim$6\farcm8$\times$6\farcm8.    
 The total integration time   was
 12240~s in  $B$, 8470~s  in  $R_c$, and 13950~s 
 in  $H_\alpha$ bands.  
 The log of the observations is given in Table~\ref{t:log}.
To complement our analysis, we retrieved from the ESO archive\footnote{
{archive.eso.org}} VLT B-band images of the same field obtained  under
Program 66.D-0069A of Gallant et al. (\cite{Gallant}) in January 2001.
These unpublished data were taken with the same telescope unit and instrument
in a high resolution mode
with an image scale of   $\sim0$\farcs1/$\mathrm{pixel}$ and
$\sim$3\farcm4$\times$3\farcm4 FOV.  The total
integration time was  about 15~ks.  
The observational log
is given in Table~\ref{t:logarchive}. 
%%%%%%%%%%%%%%%%%%%%%%%%%%%
\begin{table*}[tbh]
\caption{The same as in Table~\ref{t:log} 
but for  the 2001 period}
\begin{tabular}{ccccccc|ccccccc}
\hline\hline
 Date       &       UT  &Band, exp. & Exp.&
 secz& Seeing& Sky      & Date      &       UT 
 &Band, exp. &
 Exp.& secz& Seeing & Sky\\ 
  UT        &           & number    &  $sec.$   &
              &$arcsec.$& $\frac{counts}{sec\ pix}$ & UT      
   &           & number    &  $sec.$   &        &$arcsec.$ & $\frac{counts}{sec\ pix}$\\
\hline 
18 Jan 2001 & 07:41     & B 1   & 1000    & 1.34   &  0.51& 1.63&  -''- & 07:58	    & B 10	& 1000     & 1.33   &  0.83      & 0.51 \\
 -''-       & 07:59     & B 2  & 1000     & 1.32   &  0.57& 1.63&  -''-	& 08:16     & B 11	     & 1000	& 1.32   &  0.92 & 0.53\\
 -''-       & 08:16     & B 3  & 1000     & 1.31   &  0.64& 1.66& 25 Jan 2001 & 06:56     & B 	12     & 1000	& 1.37   &  0.58 & 0.61 \\
 -''-       & 08:38     & B 4  & 2        & 1.32   &  0.65& 6.5 &  -''-	& 07:14     & B 13	     & 1000	& 1.34   &  0.63 & 0.60\\
 -''-       & 08:39     & B 5  & 20       & 1.33   &  0.65& 6.9&  -''-	& 07:31     & B 14	     & 1000	& 1.32   &  0.67 & 0.59\\ 
19 Jan 2001 & 08:09     & B 6  & 1000     & 1.32   &  0.79& 0.96&  -''-	& 07:50     & B 15	     & 1000	& 1.31   &  0.79 & 0.58\\
 -''-       & 08:27     & B 7  & 1000     & 1.31   &  0.85& 0.97&  -''-	& 08:10     & B 16	     & 1000	& 1.32   &  0.66 & 0.57\\ 
 -''-       & 08:44     & B 8  & 1000     & 1.31   &  0.85& 1.30&  -''-	 & 08:28    & B	17           & 1000	 & 1.33   & 0.66 & 0.57\\
22 Jan 2001 & 07:41     & B 9  & 1000     & 1.32   &  0.81& 0.51 &&&&&& \\
\hline
\end{tabular}
\label{t:logarchive}
\end{table*}
 Generally, all the observations were made under photometric conditions,
instrumental magnitude variations of field stars through dataset being
insignificant, but the sky background level, seeing, and atmospheric extinction
differ.
 For instance,  the individual  images in the $B$ and 
$R_c$ bands  for { the  2003--2004 period} can be
subdivided into  two distinct groups
with a high and  low  background level  
(Table~\ref{t:log}).  The $H_\alpha$
data are more uniform in this respect,  
but seeing conditions vary  from 0\farcs67 to 1\farcs11.      
 The 2001 data were obtained under better and less variable seeing
conditions,  but the sky background level was high for the exposures taken on
January 18 and 19, 2001 during moonlight time.
(Table \ref{t:logarchive})
The data on January 25, 2001 were obtained 
 under excellent
seeing 
and sky background conditions, but  the atmospheric extinction  
was  by $\sim$0.16 magnitude higher than  the 
standard VLT value,  which may be caused by some cirri.

\begin{figure*}[t]
\setlength{\unitlength}{1mm}
\resizebox{15cm}{!}{
\begin{picture}(110,120)(0,0)
\put (0,0)   {\includegraphics[width=125mm,bb=40 150 575 655, clip]{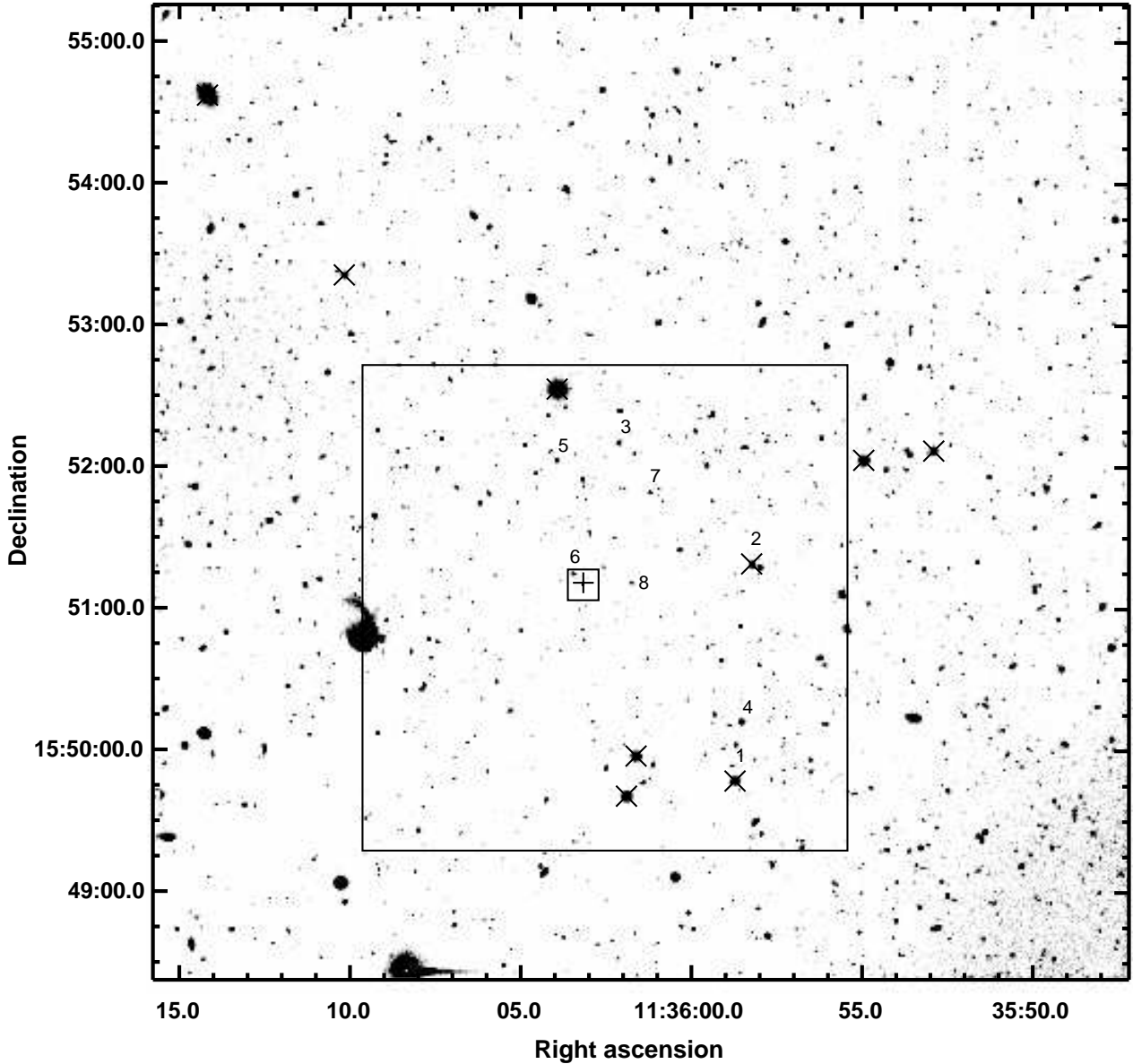}}
\end{picture}}
\caption{PSR B1133+16 field as  seen  in the $B$ band with
the VLT/FORS1 using a standard resolution mode during the 2003-2004 period. 
The position of the pulsar is marked by $+$.
  The large box marks the FOV of   
high-resolution mode observations during  the 2001 period.   
The $\times -$shaped crosses and numbers mark the USNO-B1.0 reference stars 
from Table~\ref{t:astr} and secondary photometric standard stars 
from Table~\ref{t:stars}, respectively.
The region within a small box is enlarged
 in Figs.~\ref{f:fig2} and~\ref{f:fig3}.  }
\label{f:fig1}
\end{figure*}

The data reduction 
including bias subtraction, flat-fielding, removing cosmic ray traces 
and alignment of each individual image to a reference frame was performed 
 for all the data using standard 
{\tt IRAF} and {\tt MIDAS } tools. 
 Taking the data non-uniformity into account, we considered different
combinations of individual exposures in each band to get co-added images
of the best quality, deepness, and spatial resolution.
A simple sum of all available images for a given band and period
was finally chosen as optimal and subsequently analyzed.
Only  very short, $\la30$~sec, exposures in the $B$ band
(see Tables~\ref{t:log} and \ref{t:logarchive})   
were excluded. 
As a result,  the FWHM of  a stellar object
in the composed images  was $\approx$0\farcs75~ 
and  $\approx$1\asec~ in the $B$ band  for { the 
 2001  and 2003-2004 periods}, respectively.
The respective values in the $R_c$ and $H_{\alpha}$ bands 
were  $\approx$0\farcs85 and $\approx$1\asec.  
The panoramic view
of the pulsar field as  seen
with the VLT in the $B$ band is shown in
Fig.~\ref{f:fig1}.  
%%%%%%%%%%%%%%%%%%%%%%%%%%% SECTION %%%%%%%%%%%%%%%%%%%%%%%%%%%%%%%%%%%%%%%
%%%%%%%%%%%%%%%%%%%%%%%%%%%%% Fig 1 %%%%%%%%%%%%%%%%%%%%%
%%%%%%%%%%%%%%%%%%%%%%%%%%%%%%%%%%%%%%%%%%
%%%%%%%%%%%%%%%%%%%%%%%%%%%%%%%%%%%%%%%%%%%%
\section{Astrometry and photometry} 
\label{sec3}
%%%%%%%%%%%%%%%%%%%%%%%%%%%%%%%%%%%%%%%%%%%%%
\subsection{Astrometry} 
%%%%%%%%%%%%%%%%%%%%%%%%%%%%%%%%%%%%%%%%%%%%%
%%%%%%%%%%%%%%%%%%%%%%%%%
\begin{table}[t]
\caption{The list of USNO-B1.0 stars used for astrometrical referencing with coordinates (epoch J2000.0) and their errors. }
\begin{tabular}{lcccc}
\hline\hline
 Name &  RA  & DEC &$\sigma$ RA & $\sigma$ DEC \\
      & hh mm ss &     dd mm ss& mas & mas \\ \hline
1058-0205218 & 11 35 52.8980 & +15 52 07.000 & 192 & 144 \\
1058-0205222 & 11 35 54.9493 & +15 52 03.040 &  25 &  23 \\
1058-0205225 & 11 35 58.2340 & +15 51 19.070 & 125 & 239 \\
1058-0205226 & 11 35 58.7207 & +15 49 47.110 & 307 & 138 \\
1058-0205233 & 11 36 01.6207 & +15 49 57.610 &  98 &  97 \\
1058-0205234 & 11 36 01.9007 & +15 49 40.620 &  46 &  47 \\
1058-0205238 & 11 36 03.9327 & +15 52 32.970 &  34 &  54 \\
1058-0205246 & 11 36 10.1713 & +15 53 21.120 &   4 &  19 \\
1059-0203095 & 11 36 14.1813 & +15 54 37.130 &  74 & 109 \\ \hline
\end{tabular}
\label{t:astr}
\end{table}

{  The $B$-band image obtained by
co-adding all $B$ band frames taken during the 2003-2004 period was chosen as a base image
for our astrometry because its FOV  is twice bigger than in the 2001 period.
To compute a precise astrometric image solution, the positions of objects
selected from the USNO-B1 astrometric catalog\footnote{USNO-B1 is currently
incorporated into the Naval Observatory Merged Astrometric Dataset
(NOMAD), which combines astrometric and
photometric information of Hipparcos, Tycho-2, UCAC, Yellow-Blue6, USNO-B,
and the 2MASS, www.nofs.navy.mil/data/fchpix/}
were used as a reference. There are  about 50 USNO-B1 objects in the FOV
in contrast to none from UCAC2. The recent release of the Guide Star Catalog
(GSC-II v2.3.2) \footnote{www-gsss.stsci.edu/Catalogs/GSC/GSC2/GSC2.htm}
contains a similar number of stars in this FOV but provides no information on
proper motions, and the declared astrometric errors (0\farcs3) are
higher than nominal 0\farcs2 uncertainty of USNO-B1.
We discarded the reference stars with significant proper motions and catalog-positional 
uncertainties $\ga 0\farcs4$ along with saturated ones. The pixel
coordinates of 30 objects
considered to be suitable astrometric reference points were computed making
use of the IRAF task {\it imcenter}.
The IRAF tasks {\sl ccmap/cctran} were applied for the astrometric 
transformation of the image.  Formal 
{\sl rms} uncertainties of the initial astrometric
fit were rather large,
 $\Delta$ RA = 0\farcs287 and $\Delta$ DEC=0\farcs251,
as compared with the  0\farcs2 pixel
 scale of the image.  
 Using standards with the smallest catalog
uncertainties and fit residuals significantly improved the fit.
For  nine reference stars  listed
in Table~\ref{t:astr} and marked 
in Fig.~\ref{f:fig1},  the  formal {\sl
rms} are   $\Delta$ RA = 0\farcs051 and $\Delta$
DEC=0\farcs097, which is smaller than the
pixel scale.  
As can be derived
from the  Table~\ref{t:astr},  the  local USNO-B1 catalog {\tt rms}
uncertainties of the  list in RA and DEC are  $\approx$0\farcs045 and
$\approx$0\farcs039, respectively, which is a factor of four  smaller
than the  nominal  catalog  uncertainty. To refine the fit, we further
removed standards with the largest catalog uncertainties, i.e. the first and
the fourth ones from Table~\ref{t:astr}.
This  provided much smaller {\tt rms},  $\Delta$ RA = 0\farcs0278
and $\Delta$ DEC=0\farcs068, and individual standard residuals $\la0\farcs09$,
which is consistent within 2-$\sigma$
with the local {\sl rms} of the
catalog and with the positional
uncertainties of the standards in the image  ($\la 0\farcs05$).     
  We adopted this fit as a final one.
It is important 
that by
starting with 13 standards 
the image transformation became very stable  
and practically  
independent
of  removing  other  standards
with the largest residuals. At these steps 
coordinate reference  points clustered  together 
only within  the same image pixel,
gradually migrating to some  limiting position within this pixel.  
 The same  result was obtained using the list of stars with coordinates from GSC-II v2.3.2.
Therefore, we
are quite confident 
the derived coordinate reference that is reliable 
 at least at the level of  the nominal mean catalog 
accuracy of $\approx$0\farcs2.  Combining that with the best
fit {\tt rms}, we obtained secure 1-$\sigma$ uncertainties
of our astrometric referencing  in the RA and
DEC as   0\farcs202 and  0\farcs211,
respectively.  They are comparable with the pixel scale and
 much smaller than the seeing value. } 

The co-added 2001 B-band image was rebinned by 2$\times$2 pixels to get
the same pixel scale as for the 2003-2004 period.
 All co-added 2001 $B$, $R_c$, and $H_\alpha$ images  were
aligned  to the base one using a set of  suitable unsaturated stars
with an accuracy of  better than 0.025 pixel size, and
  the above coordinate reference  of the base image was adopted for these
images.
Formal errors of this referencing  are 
negligible ($\la$0\farcs005)  in comparison with the astrometric
referencing uncertainties  of the base image.   
The selection of another summed image,  $R_c$ or $H_\alpha$,  
as a base  does not change  the result.
%%%%%%%%%%%%%%%%%%%%%
\begin{table*}[th]
\caption{\object{\psr}  radio 
coordinates at the epochs of
the VLT and Chandra observations. }
\begin{tabular}{lcccc}
\hline\hline
 Epoch &  RA  & DEC  &$\sigma_{RA}$ & $\sigma_{DEC}$                                \\
2450000+      & hh mm ss &     dd mm ss& mas & mas                                          \\ \hline
1544.5/2000$^*$       & 11 36 03.1829 & +15 51 09.7257 & 15  & 15                \\
1932.5/2001 Jan. & 11 36 03.1774 & +15 51 10.1189 & 15  & 15                      \\ 
2738.5/2003 Apr  & 11 36 03.1662 & +15 51 10.9228 & 15 & 15                        \\
2997.5/2003 Dec. & 11 36 03.1625 & +15 51 11.1898 & 15 & 15                       \\ 
3029.5/2004 Jan. & 11 36 03.1621 & +15 51 11.2221 & 16 & 15                        \\ 
3424.5/2005 Feb. & 11 36 03.1565 & +15 51 11.6201 & 16 & 16                        \\ 
\hline 
\end{tabular} \\
\label{t:psrcor}
\begin{tabular}{l}
$^*$ Brisken et al.~\cite{Brisken} 
\end{tabular}
\end{table*}
%%%%%%%%%%%%%%%%%%%
    
The  radio positions  of  \object{\psr}  at the  epochs  of the VLT and Chandra observations 
(Table~\ref{t:psrcor}) were determined using  radio measurements of
the pulsar proper motion made with the VLA by Brisken et al. \cite{Brisken}. 
The position errors in Table~\ref{t:psrcor}, 
$\sigma_{RA}$ and $\sigma_{DEC}$, include the errors
of the radio position at the reference epoch
J2000  (15 mas for both coordinates) and the uncertainties in the pulsar proper motion 
(0.38 mas and 0.28 mas in RA and Dec, respectively).
%%%%%%%%%%%%%%%%%%%%%

 Combining 
the  errors of  our astrometry 
and the radio position errors,   
 we derived  uncertainties  of
the pulsar position in our images.
Observations in different bands were splinted by different sets
on a time basis from a week to 8-9 months, and
the pulsar has a large
proper motion, particularly along the declination (Table~\ref{T.1})
(cf.~Tables~\ref{t:log} and \ref{t:logarchive}), 
therefore one has to account for a systematic uncertainty related to
a shift in the pulsar position during this
time. It is mostly significant
for the B and R band observations of the
2003-2004 period  when the pulsar shifts 
by  $\sim$0\farcs197 and $\sim$0\farcs229, or by about 1 CCD pixel,  
from the beginning to the end of the
observations. { We added these shifts 
as uncertainties squared and calculated the
expected optical counterpart coordinates as a
simple mean of those at the beginning and the end of
the observations  for  given resulting frames
and  periods. 
The results are summarized in Table~\ref{t:optcor}.
As seen,  despite  a rather accurate astrometric
referencing of the summed 
images,  the  resulting  $3\sigma$ position
uncertainty of the expected optical counterpart  
can be as large as  1 arcsecond in the DEC due to the large
proper motion of the pulsar.}      

%%%%%%%%%%%%%%%%%%%%%%%%%%%%%%%%%%%%%%%%%%%%%%%%%%
\subsection{Photometric calibration of the B and R$_c$ images}                
%%%%%%%%%%%%%%%%%%%%%%%%%%%%%%%%%%%%%%%%%%%%%%%%%%%%%%

For precise photometric calibration of all  datasets taken in different nights and periods,
a two-step approach known as ``differential photometry'' was applied.
At first, absolute magnitudes
of a   set of  relatively bright stars visible in all target frames (so
called ``secondary standards'') were derived accurately. This was done using
images of the primary Landolt's standards obtained during the same night
chosen as a  ``reference''.   
 In the  second step   the  co-added target images were directly calibrated
 using the  secondary standards.  
 An advantage of this approach is that it is not necessary
 to consider possible magnitude zero-points and/or extinction variations 
 from night to night since the secondary 
 standards and the target  are in the same frames. 
%%%%%%%%%%%%%%%%%%%%%%%%%%%%%%%%%%%%%%%%%%
\begin{table}[t]
\caption{ Expected optical counterpart coordinates   
 at the summed VLT frames. }
\begin{tabular}{lcccc}
\hline\hline
 Frame/   &  RA  & DEC  &$3\sigma_{RA}$ & $3\sigma_{DEC}$                                \\
  Period          & hh mm ss &     dd mm ss& arcsec &arcsec                                   \\ \hline
$B$/2001        &11 36 03.1774 & +15 51 10.1189
& 0.61  & 0.63                \\
$B$/2003        & 11 36 03.1644 & +15 51 11.0563
 &  0.63  & 1.02                      \\ 
$R_c$/2003-04  & 11 36 03.1642 & +15 51 11.0719
& 0.63 & 1.10                        \\
$H_\alpha$/2004 & 11 36 03.1621 & +15 51 11.221
& 0.61 & 0.63                       \\ 
\hline 
\end{tabular}
\label{t:optcor}
\begin{tabular}{l}
\end{tabular}
\end{table}
%%%%%%%%%%%%%%%%%%% 

The night of 02/03 April  2003
was used as a reference for  photometric calibration and,  for this night, the zero-points\footnote{The zero-points 
were calculated for the  flux  in ADU/s units. 
The Paranal Observatory web page 
lists this night  as a photometric one and our 
zero-points, when converted to electron/s units,  
are consistent with those
provided for this night by the  VLT/FORS1 team.}   in the Johnson-Cousins $B$ and $R_c$ bands, and color terms establishing the relation between the real $B$ and $R_c$ magnitudes 
and respective   instrumental magnitudes  
$m_B$ and $m_R$ were derived using 14\footnote{  A maximum number  of  standards  per   night  observed  during 
our program.}  Landolt's standard stars  from \object{Rubin149}, \object{PG1047+003},  and  \object{SA110} fields  (Landolt \cite{Landolt}):    
\begin{eqnarray}
B &= & 27.17(3)+m_B -0.04(2)(m_B-m_R)\\ \nonumber
R &= & 27.38(2)+m_R+0.01(2)(m_B-m_R).
\end{eqnarray}

For a star-like object, the instrumental magnitude in  
j-th band  is defined as 
\begin{equation}
m_j= -2.5~log(f^j_{ap}/t_{exp})-\delta m_j
-k_{j}~secz~,
\end{equation} 
where $f^j_{ap}$ is the source flux in counts 
for a given 
aperture, $t_{exp}$  the exposure time, $k_j$ the extinction
factor, 
$secz$ is airmass, and   $\delta m_j$  the correction 
for a finite aperture derived from a point spread function (PSF) 
of bright stars in the image. 
The signal-to-noise ratio $S/N$ and  magnitude uncertainty   
$\Delta m$ are calculated in a standard way as (Newberry \cite{newberry})  
\begin{equation}
\frac{S}{N} =
\frac{f_{ap}}{\sqrt{f_{ap}/g+n_{ap}\sigma_{bg}^2(1+1/n_{bg})}} 
\end{equation} 
\begin{equation}
\Delta m = 1.0856\ \Bigl(\frac{S}{N}\Bigr)^{-1}, 
\end{equation} 
where $\sigma_{bg}$ is the standard deviation of the background
in counts, $n_{ap}$ the number of pixels in the source
aperture, $g$  the gain, and $n_{bg}$  the number of pixels in area 
used for the background measurement. 
With our seeing conditions, the source aperture with a radius 
of 3 pixels (at the image scale $\approx$0\farcs2/$\mathrm{pixel}$), 
for which  $S/N$ reaches  a maximum,   
was found to be optimal for  stellar objects in all 
our images. 
The atmospheric transparency  during our reference night was 
close to a standard value and we used the nominal extinction
factors $k_B=0.25$ and $k_R=0.08$ provided for the VLT
site\footnote{ http://www.eso.org}. 
%%%%%%%%%%%%%%%%%%%%
\begin{table}[tbh] 
\caption{Secondary standard stars used for photometric referencing. }
\begin{tabular}{lccc|lccc}
\hline\hline
 NN &  B  & $R_c$  & $H_{\alpha}^{AB}$ &NN & B & $R_c$ & $H_{\alpha}^{AB}$  \\
    & mag &  mag   &  mag            &   & mag& mag  &  mag \\ \hline   
1   & 19.91 & 19.24 & 19.35 &5 & 22.84 & 21.40 & 21.63 \\  
2   & 21.10 & 20.23 & 20.37 &6 & 23.04 & 22.46 & 22.88 \\ 
3   & 22.07 & 20.47 & 20.57 &7 & 22.98 & 20.43 & 20.35  \\
4   & 21.27 & 20.71 & 21.52 &8 & 24.05 & 23.32 & 23.23 \\ \hline
\end{tabular}
\begin{tabular}{l}
errors are $\pm$0\fm03 in the $B$ and $R_c$, and up to $\pm$0\fm05  
in the $H_\alpha$ bands \\
\end{tabular}
\label{t:stars}
\end{table}
%%%%%%%%%%%%%%%%%%%%% 

The $B$ and $R_c$ magnitudes of the stars chosen as secondary standards
were measured at the reference-night images calibrated
with the primary standards as described above.
The stars are  marked by numbers in Fig.~\ref{f:fig1}, 
and  their magnitudes with errors accounting for the zero-point errors shown in brackets of 
 Eq (1) are listed in Table~\ref{t:stars}.

Using the secondary standard magnitudes, the resulting zero-points  obtained 
for the  images summed over the whole {  2003-2004 period}    are 
$B$=27.09$\pm$0.03 and $R_c$=27.37$\pm$0.02.   
For the 2001 period
the resulting $B$ band zero-point  
is  26.93$\pm$0.03. 
We also  estimated  $3\sigma$  
detection limits of a point-like object in the
co-added images. In the $B$ band it
is $\sim$28\fm2 for  the both   2001
and 2003-2004 periods, and $\sim$28\fm6 for the sum 
of the two periods. 
It is $\sim$27\fm9 in the $R_c$ band.
There are no significant difference between the $B$ band formal 
detection limits for both periods. However,  
an ``eye visibility'' of a $3\sigma$ object
is better for  the 2001 period.  
This is due to better seeing conditions and   
a higher FORS1 resolution  setup  during  this period (see Sect.~\ref{sec2}).  
 
 Stellar magnitudes $M_j$ 
can be transformed into  absolute fluxes $F_j$ 
(in erg~s$^{-1}$~cm$^{-2}$~Hz$^{-1}$), whenever  necessary,  
using standard equations  
\begin{equation}
\log F_j = - 0.4 \left(M_j + M_j^0\right),
\label{eq:fl_mag}
\end{equation}
with the zero-points provided by Fukugita
(\cite{Fukugita}): $M_B^0=48.490,\
 M_R^0 = 48.800$.

%%%%%%%%%%%%%%%%%%%%%%%%%%%%%%%%%%%%%%%%%%%%%%%%
\subsection{Flux calibration of the H$_{\alpha}$ image}
%%%%%%%%%%%%%%%%%%%%%%%%%%%%%%%%%%%%%%%%%%%%%%%%%%%%
Flux calibration in the H$_{\alpha}$ narrow band ($\lambda$=6563\,\AA, FWHM=60\,\AA) image
was done  using  
the  spectrophotometric standard \object{HZ44} 
observed  the same nights as the pulsar.
The flux $F_{H_\alpha}$ and
AB-magnitude
$M_{H_{\alpha}}^{AB}$  of a source in this band  
are defined as 

\begin{equation}
 F_{H_\alpha} 
  =  
 C \ {\frac{ f\times 
10^{~(0.4~k_{H_\alpha}~secz)}} {t_{exp}} \ \frac{erg}{ cm^{2}\ s\
Hz}}
\end{equation}
\begin{equation}
M_{H_{\alpha}}^{AB} = -2.5~log(F_{H_{\alpha}})-48.6, 
\end{equation}
where $f$ is the measured flux of the source  in counts and $C$ 
 the calibration constant.

The calibration was performed using the night of 19 January 2004 
as a reference. From count-rate variations of stars in the pulsar field 
with the airmass,  we derived a mean atmospheric extinction factor 
in the H$_{\alpha}$ band $k_{H_{\alpha}}=0.055\pm0.017$. 
Then  the calibration constant for this night  
was derived to be 7.66$\times$10$^{-30}$.   
The H$_{\alpha}$ magnitudes of the secondary standards  
measured using the reference night image
are listed in Table~\ref{t:stars}.  Based on
these magnitudes the derived $3\sigma$ flux detection limit for a stellar object  
in the  summed  H$_\alpha$ image   is 1.1$\times$10$^{-30}~\mathrm{erg\ cm^{-2}\ 
s^{-1}\ Hz^{-1}}$, and it is    
5.5$\times$10$^{-31}~\mathrm{erg\ cm^{-2}\ s^{-1}\ Hz^{-1} 
\ arcsec^{-2}}$ for a surface brightness of an extended object. 
%%%%%%%%%%%%%%%%%%%%%%%%%%%%%%%%%%%%%%%%
%
%%%%%%%%%%%%%%%%%%%%%%%%%%%%%%%%%%%%%%%%%%%%% Fig 2 %%%%%%%%%%%%%%%
\begin{figure*}[t]
\setlength{\unitlength}{1mm}
\resizebox{12cm}{!}{
\begin{picture}(120,165)(0,0)
\put (-01,83)   {\includegraphics[width=86.5 mm,  bb= 55 202 520 590,clip]{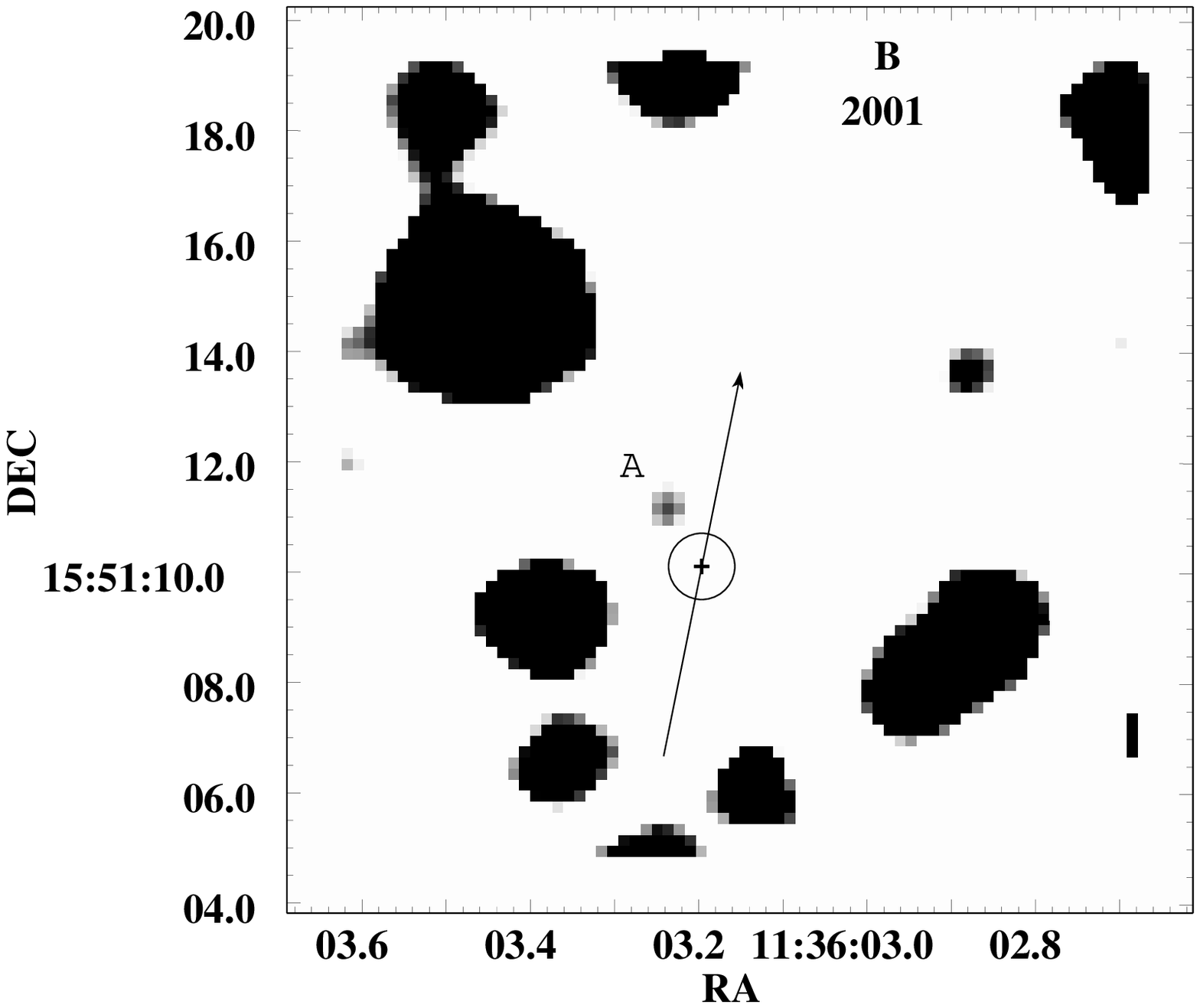}} % 2001.B.pub2.ps}}
 \put (88,83)   {\includegraphics[width=83.5 mm,  bb=60 202 506 590, clip]{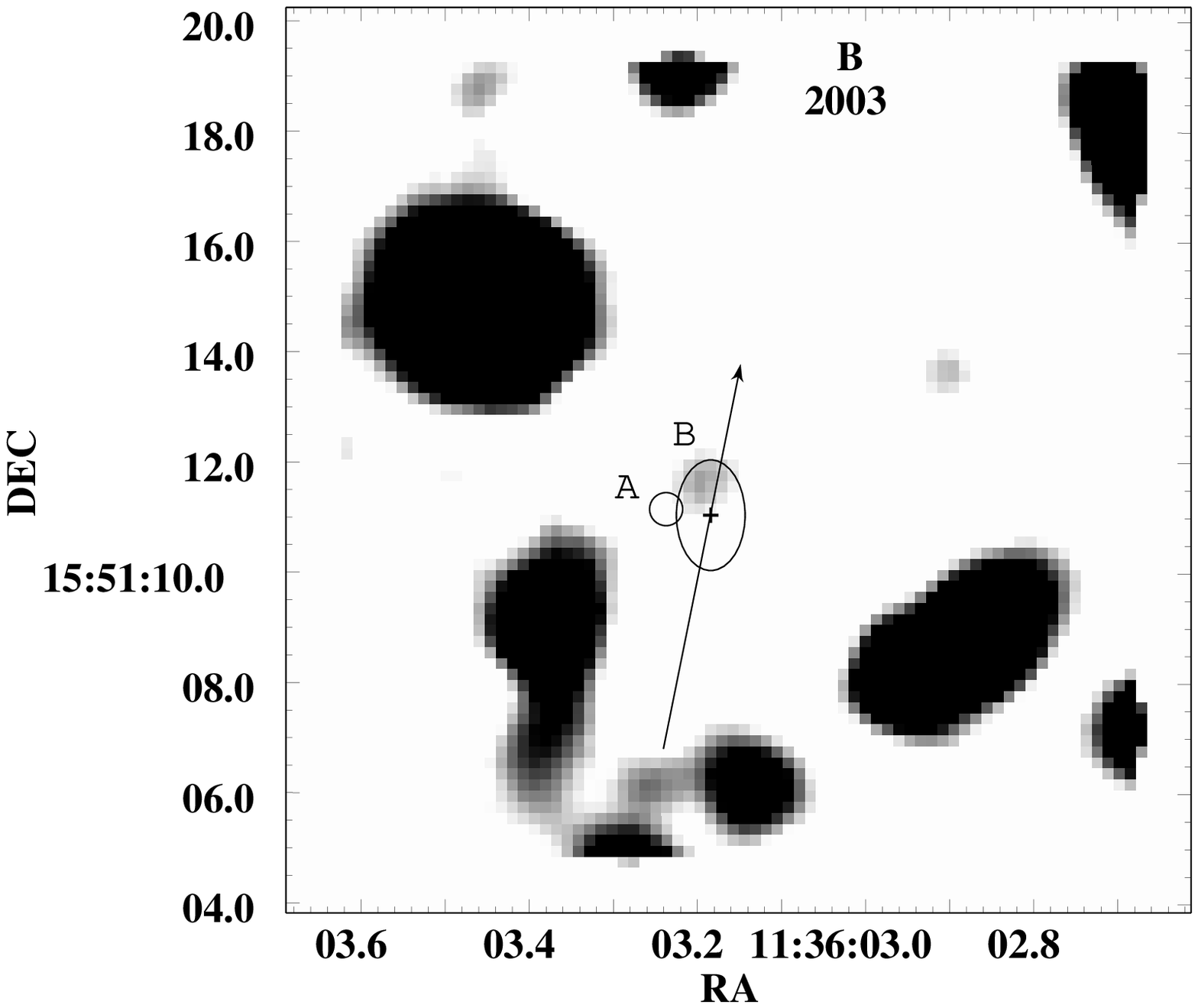}}  %2003.B.pub2.ps}}  
\put (00,00)   {\includegraphics[width=86.5 mm, bb=145 299 402 528, clip]{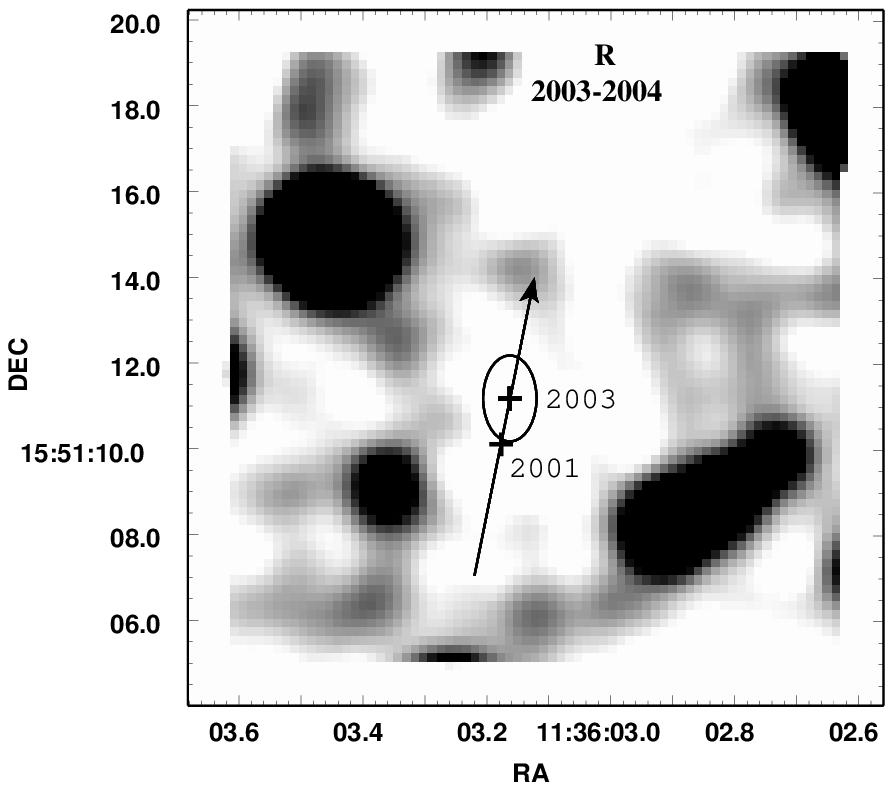}}  %zharikovfig2c.ps}}
\put (88,01.5)   {\includegraphics[width=83.5mm, bb=88 253 458 583, clip]{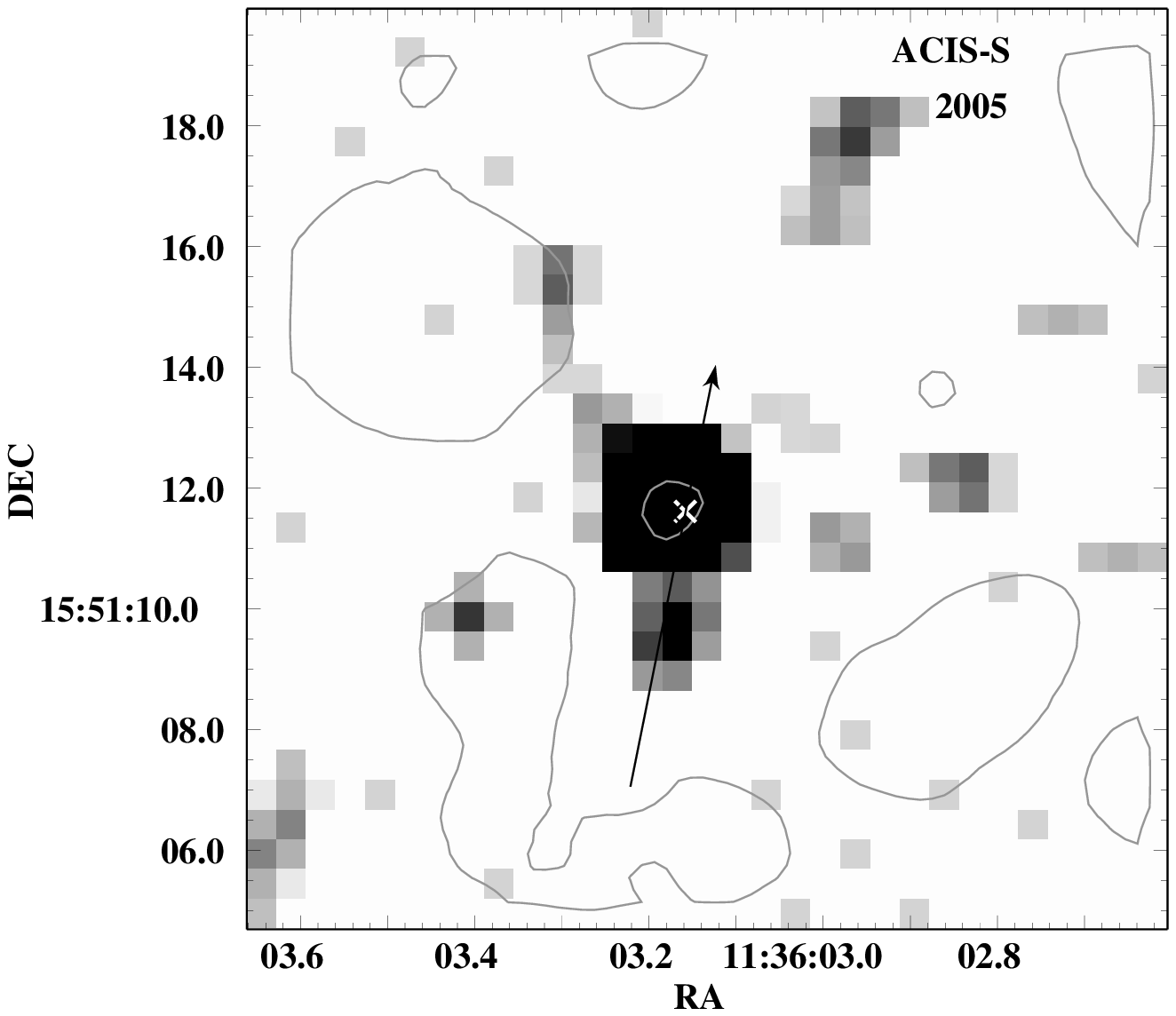}}  %zharikovfig2d.ps}}
\end{picture}
}
\caption{{ 
Fragments  of the \object{\psr} 
field images  
obtained in the optical  $B$ and $R_c$ bands with the VLT/FORS1 and
in 0.2--10 keV X-ray range with the Chandra/ACIS-S at different epochs,
as indicated in the plots.
All images are smoothed with  $\sim$1\asec~Gaussian kernel, and the fragment sizes are $\sim$14\asec$\times$14\asec. }
 The $+$ and $\times$    mark the expected radio positions of the pulsar
at the observing epochs in  the optical and X-ray images, respectively. 
The ellipses  show  
$3\sigma$ position uncertainties, 
and the arrows show the direction of the pulsar proper motion 
and its path over about 20 yrs.
The shift in the pulsar from 2001 to 2003 is indicated in the $R_c$ image. The contour 
of 
 object  $A$ seen in the 2001 $B$-band image is overlaid on the 2003 B-band image.
 The contours on the X-ray image are from the 2003 $B$-band image. Note that
 object $B$  in the 2003 $B$-band image  nicely fits  the position of the candidate
 X-ray counterpart of the pulsar in the ACIS-S image.  
}
\label{f:fig2} 
\end{figure*}

%%%%%%%%%%%%%%%%%%%%%%%%%%%%%%%%%%%%%%%%%%%%%%%%%%%%%%%%%%%%%%
%%%%%%%%%%%%%%%%%%%%%%%%%%%%%%%%%%%%%%%%%%%%%%%%%%%%%%%%%%%%%%
\section{Results}
\label{sec4}
%%%%%%%%%%%%%%%%%%%%%%%%%%%%%%%%%%%%%%%%%%%%%%%%%%%%%%%%%%%%%%
\subsection{Searching for the pulsar counterpart in the B and R bands}
Zoomed  fragments  of the field containing the pulsar in the $B$ and $R$ bands for 
 the   2001 and 2003-2004 periods 
are   
presented in  Fig.~\ref{f:fig2}.   
For comparison we also show the X-ray image of the same fragment
obtained with the Chandra/ACIS-S in February 
2005\footnote{The X-ray data were retrieved from the Chandra archive (Obs. ID 5519, 2005-02-23.18ks
exposure, PI G. Garmire.} where a candidate 
 X-ray pulsar 
counterpart has been found (Kargaltsev et al.~\cite{Karga06}).    
The expected  positions of the pulsar for each
 observing epoch,
together 
with the $3\sigma$ position uncertainty ellipses
from Tables~
\ref{t:psrcor} and \ref{t:optcor},  are marked
with the pulsar proper motion direction.

There are no significant objects within the pulsar position error ellipse in
the 2001 $B$-band image.
The nearest detected object, marked  $A$, is  outside the ellipse and  { lies
 1\farcs1 away} from the pulsar 2001 position in a direction not coinciding 
with  the pulsar proper motion.  { Its $B$ magnitude is  $28.0\pm0.3$ or only 
about $3.5\sigma$ of the  formal detection limit.} The outer contour of this object, 
where it merges with backgrounds, is overlaid on the 2003 $B$-band
image.  As seen, it only  partially overlaps
the pulsar error ellipse  at the epoch of 2003, 
although its center is very near to the ellipse border.

At the same time, within the pulsar position uncertainty ellipse in the 2003 $B$-band image,
we find another faint object  ($B$)
with a similar magnitude   $B =28.1\pm0.3$. 
By a good  positional coincidence with the
expected pulsar coordinates, the object $B$
can be considered as a candidate   pulsar
optical counterpart.  The object  $A$ also  can be  resolved    
 in this image within its contour of the 2001 epoch (see below).  
 However, it sits in
the east wing of the candidate   
and is not visible  in the 2003 $B$-band
image  of  Fig.~\ref{f:fig2}, where
the  count range is chosen in a such way as to underline the
presence of object  $B$.   The positions and magnitudes
of objects $A$ and $B$ are listed in Table~\ref{t:psrcount}.

A thorough inspection of the region containing
  objects $A$ and $B$  in the 2001
and 2003 $B$-band images was made with using
individual exposures.  It confirmed  
that  both objects are real
but not artifacts  caused by,  e.g., a poor cosmic ray removing
or flat-fielding.  
Owing to their closeness  and faintness, 
these objects   may represent  bright parts of   the same unresolved   extended 
background feature.  To verify this possibility  we considered   
the changes in spatial profiles of the region containing
both objects from  one observing epoch to
another.     Within positional uncertainties,  the objects  almost have  
the same declinations,  while their right ascensions  differ by about
one arcsecond.  This allows us to  compare only  
1D E-W spatial   profiles   
at a fixed declination. 

\begin{figure}[t]
\setlength{\unitlength}{1mm}
\resizebox{6.5cm}{!}{
\begin{picture}(95,125)(0,0)
\put (0,10)   {\includegraphics[width=125mm, bb=30 120 392 447, clip=]{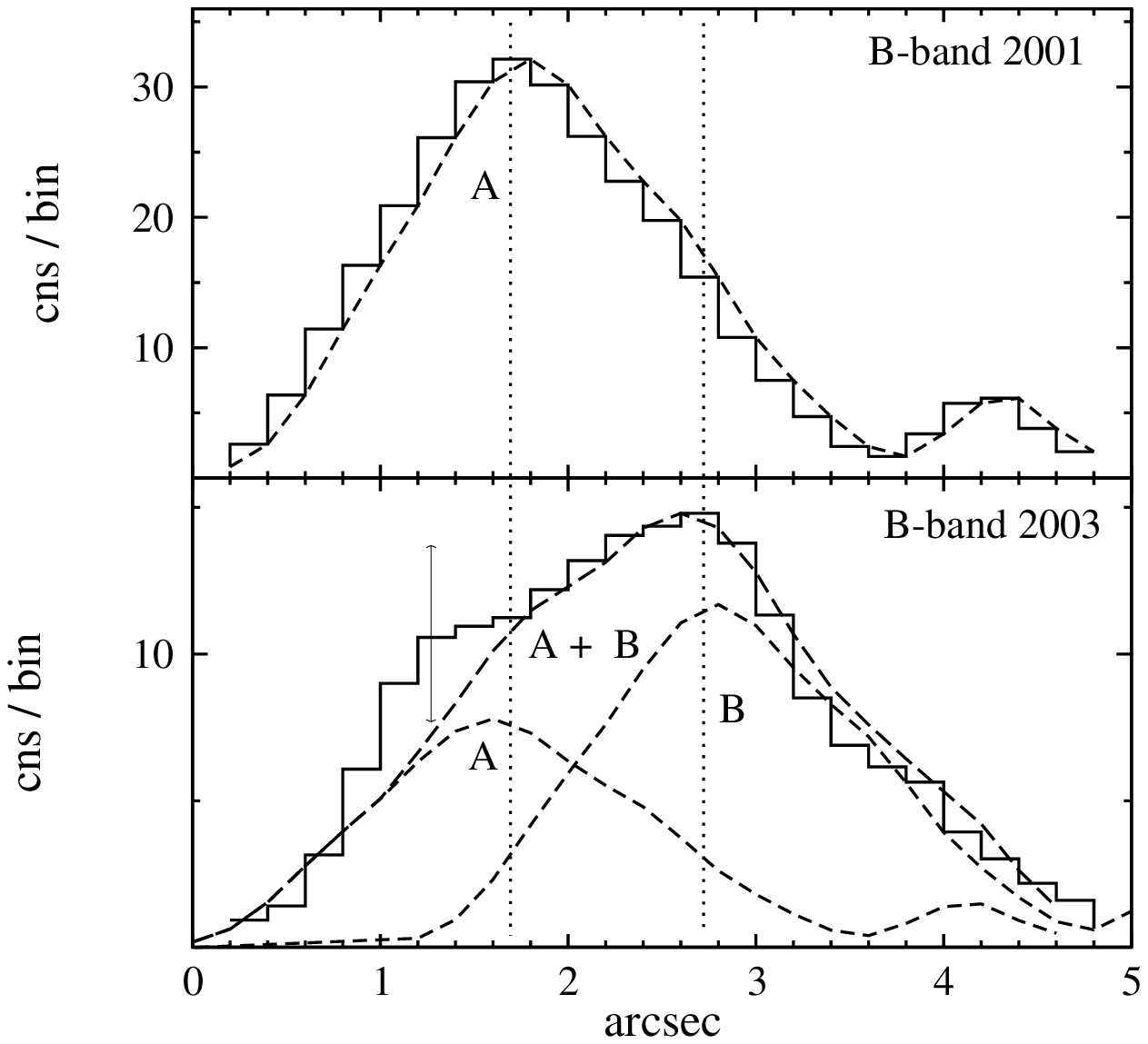}} %ab-prof-v2.eps}}
\end{picture}
}
\caption{The observed E-W spatial profiles of the  regions
containing  objects $A$ and $B$
extracted  from the 2001 (top) and 2003 (bottom)
B-band images  of Fig.~\ref{f:fig2}  along   the   slit
with  PA of 90$^o$  and 
Dec=+15:51:11.49.  
The slit length is 4\farcs6 and  its  width 
is 0\farcs6. 
The coordinate origin of the horizontal axis corresponds 
to the eastern edge of the slit  with  
RA=11:36:03.317. The background was subtracted. 
 Vertical dots  mark the 
 positions of  
$A$ and $B$  defined from  the  peaks  of
the profiles in the top and bottom panels,
respectively.  The short dashed line in the top panel shows 
a smoothed  $A$ profile that is used as a template to fit
the $A +B$  profile in the bottom panel as a sum of the two
sources. The resulting fit is shown by the  long dashed line, 
while short dashed lines show the contributions of $A$ and $B$.
}
\label{f:figAB} 
\end{figure}

In Fig.~\ref{f:figAB}  we present 
the 1D profiles   extracted from the 2001 and
2003 $B$-band images, shown in Fig.~\ref{f:fig2},
along a horizontal slit 
with PA=90$^o$ and DEC=+15:51:11.49.    
The slit length and width 
are 4\farcs6 and 0\farcs6, respectively, and the centers  of 
both objects  $A$ and $B$  are within this slit.
 The coordinate origin of the horizontal axis in Fig.~\ref{f:figAB} corresponds 
to the eastern edge of the slit  with RA=11:36:03.317.  
 Only   object $A$ is seen  in
 the 2001  profile  ({\sl top panel}).
 The 2003  profile  is significantly wider, about twice,  and has
 an asymmetric shape  
 suggesting two sources of
different intensities with overlapping
profiles ({\sl bottom panel}). Its peak is shifted towards west  
and corresponds to the position of   object $B$, while
 object  $A$  is probably responsible for
the  ``shoulder''  in the east profile wing.

The objects are too faint  to be resolved reliably as two point-like
sources with standard PSF modeling and subtraction
tools.    To analyze the situation,  
we applied   a simplified approach based on the fact that  
only  a single object $A$ is resolved  in  the 2001 $B$-band image and profile.
We used its profile shape  as a template of a single source profile
to fit   the  suggested    $A+B$  blend  at the 2003 period
by a sum of the emissions from two sources, assuming  
that $B$ has the similar profile shape.
  Within  uncertainties,  this model   is in a good agreement with the 
2003 observations 
if     the  peak intensity of  object $A$  is  by a factor of
1.5 lower than that of $B$. This corresponds to about  a 0.4 magnitude
difference,  which is within the uncertainties of  the objects'
magnitude measurements.     
The distance between $A$ and $B$ is about
0\farcs8-1\farcs0.  This  simplified approach supports the 
interpretation  that  $A$ does not change  its position and
brightness significantly from the 2001 epoch to 2003, while $B$ appears
in the 2003 image as an additional source  roughly of the
same brightness, and  by positional coincidence it can be  associated with the 
pulsar.     We discuss this and other possibilities of the $A+B$
interpretation  in Sect.  \ref{sec5}.      
    
 %%%%%%%%%%%%%%%%%%%%%%%%%%%%%% Fig 3 %%%%%%%%%%%
%%%%%%%%%%%%%%%%%%%%%%        
%%%%%%%%%%%% Table 7 %%%%%%%%%%%%%%%%
\begin{table*}[t]
\caption{The parameters of the optical objects detected around the expected position of \object{\psr} }
\begin{tabular}{lcccl}
\hline\hline
 Object &  RA  & DEC  &epoch & magnitude                                    \\
  label        & hh mm ss &     dd mm ss&  &                                       \\ \hline
A    & 11 36 03.22$\pm$0.04 & +15 51 11.16$\pm$0.60 & 2001 &B$=28.0\pm0.3$  \\
B    & 11 36 03.16$\pm$0.04 & +15 51 11.49$\pm$0.60 & 2003 &B$=28.1\pm0.3$,   $B-R<0.5$ \\ 
C    & 11 36 03.15$\pm$0.04 & +15 51 10.86$\pm$0.60 & 2004 & $H_\alpha^{AB}=26.3\pm0.4$  \\
\hline
\end{tabular}
\label{t:psrcount}
\end{table*}

Neither  of the  objects $A$ and  $B$ are resolved
 in the $R$-band image, which is obviously shallower due to a shorter integration
 time (cf.~Sect 2). 
Nevertheless, the absence of a ``red'' source at the expected pulsar positions 
allows us to constrain the color index for the candidate detected in the $B$ band as
$B-R \lesssim0.5$.

To better compare the 2005  X-ray  and the 2003  $B$-band images we overlaid the contours
of the optical image onto the X-ray image. The pixel scale of the ACIS-S image is 
about 0\farcs5, which is comparable to the proper motion shift of the pulsar between 
the 2003 and 2005 epochs, but about
three times smaller  than the FWHM of the ACIS PSF. Within these uncertainties    and a nominal $\sim$1\asec~Chandra/ACIS pointing accuracy, 
the contour of the possible optical counterpart,
object $B$,  fits the position
of the candidate 
X-ray counterpart  nicely, which is the brightest object in the X-ray image.  
This implies, at least, that we  probably see the same object in the optical and X-rays. 
The position of the X-ray candidate with 0\farcs2 accuracy coincides with the 
expected  radio position of the pulsar at the epoch of the Chandra observations 
(Kargaltsev et al.~2006).
We also note that counterparts of some other optical objects in the field may 
be found in the X-ray image and {\it vice versa}.           

%%%%%%%%%%%%%%%%%%%%%%%%%%%%%%%%%%%%%%%%%%%%%%%%%
\subsection{Possible pulsar tail in $H_\alpha$  and X-rays?}

%%%%%%%%%%%%%%%%%%%%%%%%%%%%%%%%%%%%%%%%%%%%%%%%%%%
 %%%%%%%%%%%%%%%%%%%%%%%% Fig 4 %%%%%%%%%%%%%%%%%%%%%%%%%%%% 
The 2004 $H_\alpha$-image of the same field and the difference in the $H_\alpha$ and  normalized $R$-band
images, where  most of the continuum emission is subtracted,   are shown
in Fig.~\ref{f:fig3}. The contours of the $\mathrm{ACIS-S}$  X-ray image, which is also shown  here but 
with a slightly different gray scale 
 than in Fig.~\ref{f:fig2}, 
are overlaid on the ($H_\alpha-R$)-image and vice versa. As in Fig.~\ref{f:fig2}, all images
are smoothed with the Gaussian kernel of about 1\asec,  which roughly corresponds
to the seeing  of the resulting optical images.   

In the $H_\alpha$ image we find a faint point-like source ($C$), which { lies} within  the pulsar
position error ellipse. It is near the detection limit and its star-like $H^{AB}_\alpha$ magnitude 
is 26.3$\pm$0.4 ($2.7\sigma$ formal detection limit), which corresponds to the flux 
$F_{H_{\alpha}}\approx 1.05\times 10^{-30}$ ergs cm$^{-2}$ s$^{-1}$
Hz$^{-1}$ or $1.5\times10^{-6}$ photon cm$^{-2}$ s$^{-1}$.  
 The object is not detected in the  $B$ and $R_c$ bands, while  
its possible artifact origin 
in the   narrow band was ruled out  
after a careful inspection of  individual $H_\alpha$ frames.

 We do not see any characteristic extended bow shock 
structure around the pulsar, as  is produced, for instance, around another high-velocity pulsar 
\object{PSR B2224+65}, known as the Guitar nebula (Cordes et al.~\cite{cordes}). Instead of that, we see a faint and rather 
clumpy $H_\alpha$ emission  most likely associated with the recombination of a heated ambient matter around 
background  objects of the field. Some of them may have X-ray counterparts. 
However, there is one faint  structure (about 4.5\asec~length) south of
 object $C$. 
 This apparent structure is roughly aligned  with 
the pulsar proper motion direction and located behind its 2003  position.
It is better seen in the (H$_\alpha$-R) image where it merges
with 
object $C$. There are no signs of the structure in the broad band optical images. At the same time,
in the X-ray image,  we  also see   a marginal tail-like structure behind the pulsar counterpart 
candidate, though of smaller spatial extent.  It was not  mentioned  by Kargaltsev et al. \cite{Karga06}, who only reported 
 on the detection of the X-ray pulsar candidate.
 If  both 
 structures are  not  background fluctuations,  they can be considered as 
cometary-like pulsar tail candidates. 

%%%%%%%%%%%%%%%%%%%%%%%%%%%%%%%%%%%%%%%%%%%%%%%%%%%%%%%%%%%%%%%
\section{Discussion} %Optical emission of \psr}
\label{sec5}
%%%%%%%%%%%%%%%%%%%%%%%%%%%%%%%%%%%%%%%%%%%%%%%%%%%%%%%%%%%%%%%%
%%%%%
In Table \ref{t:psrcount} we collected the coordinates
and magnitudes  of the objects detected around the radio positions of the pulsar in each VLT observing epoch. % of observations.
\begin{figure*}[t]
\setlength{\unitlength}{1mm}
\resizebox{12cm}{!}{
\begin{picture}(120,165)(0,0)
\put (00,83)   {\includegraphics[width=84.5mm, bb=144 296 400 524, clip]{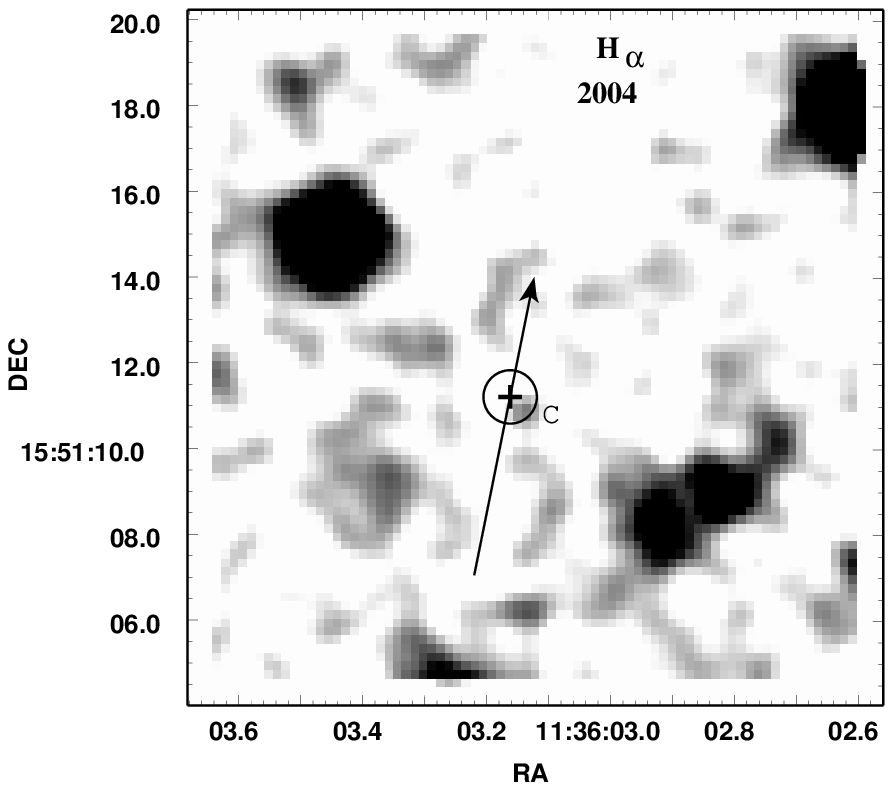}}
\put (86,78)   {\includegraphics[width=84.5mm, bb=147 296 404 539, clip]{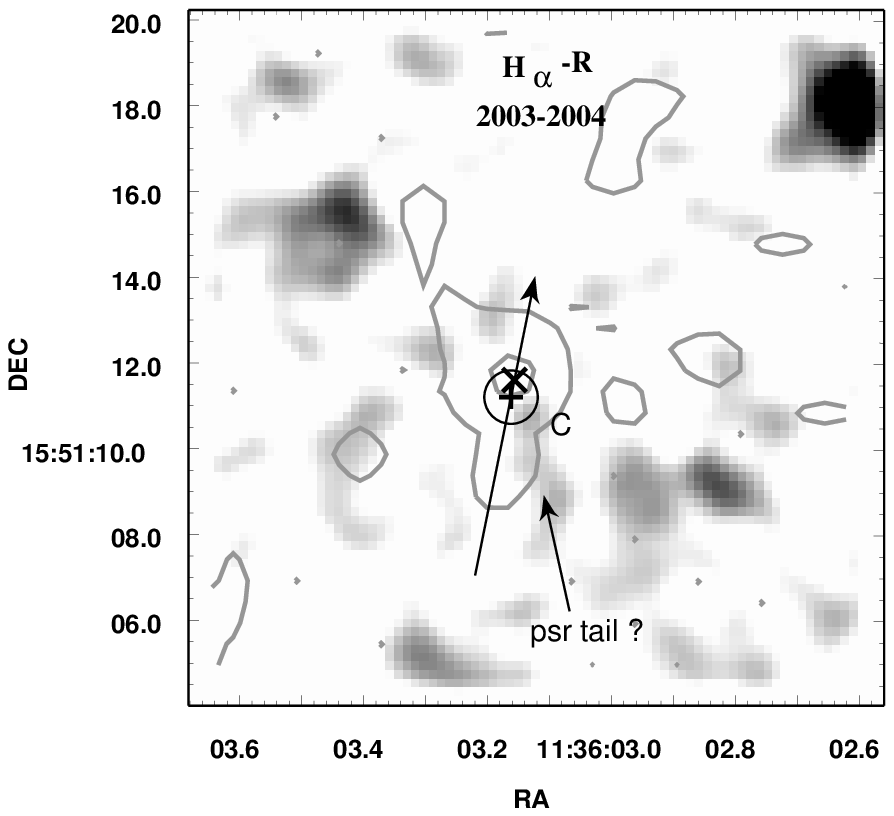}}
\put (86,00)   {\includegraphics[width=84.5mm, bb=73 245 462 584, clip]{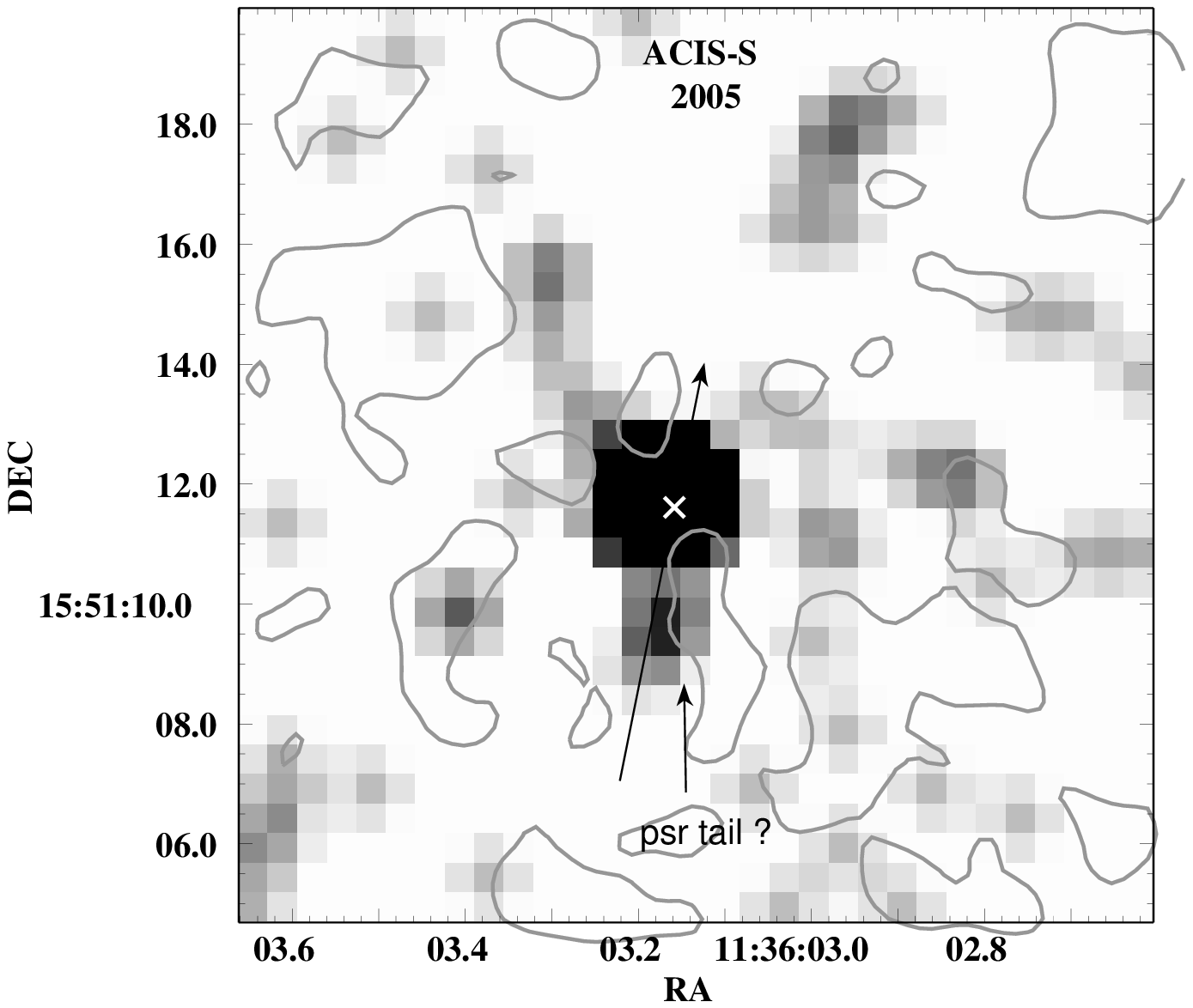}}
\end{picture}
}
\caption{The same as in Fig.~\ref{f:fig2} but for the  H$_\alpha$
and (H$_\alpha$-R) images,
with the contours from the X-ray image overlaid.
 The X-ray image with the (H$_\alpha$-R) image contours
overlaid is also shown for comparison.
      }
\label{f:fig3} 
\end{figure*}
%%%%%%%%%%%%%%%%%% end Fig 4
  Given that   sources $B$ and $C$
are located 
within   3$\sigma$ pulsar position error ellipces related  
to respective observing epochs  (cf., Tables~\ref{t:optcor} and
\ref{t:psrcount}),   
they can be  associated with the pulsar. 
Object  $B$,
emitting in a broadband range,
can be considered as a candidate
pulsar optical counterpart, while  $C$ can be
a  sign  of the pulsar  interaction with a clumpy ambient  ISM  seen via the H$_{\alpha}$ 
recombination line.      
  Object  $A$ is outside of the  $3\sigma$-error ellipses for both  
observing  epochs  and we qualify it   as a background  source. 
If  object $B$ detected in the 2003 B-band image  is the
real counterpart of the pulsar, 
then why  do  we not  resolve it in the expected place on the 2001 $B$  image? 

One of the reasons is the faintness of the object, which is resolved at a level very 
close to the estimated 3-$\sigma$ detection limits (see Sect. 3.2). 
  To estimate the probability of detecting an  object with  S/N $\sim3$ in our data, 
we selected almost  a homogeneous subset of 2001 B-band images, B12-B17,  which provides
a maximal number of frames with  the same seeing, background, and atmospheric extinction 
(see Table ~\ref{t:logarchive}). Then, we  selected about  100 faint objects  chosen 
randomly  in the sum of these 6 frames and checked that these objects are 
detected in the each of the individual frames. That was done in  two ways. 
In the first,  we used the optimal aperture, with 3 pixel radius, and an object 
was accepted as  detected at an individual image if its magnitude error was 
less that 0.36, which corresponded to $S/N\approx 3$.  In the second way we estimated 
the object by eye  in the  images  smoothed  with the Gaussian kernel.
In the first case the probability 
of detecting  an object with $S/N\sim 3$ was 60\% -- 70\%.  In the second, less formal case,  
it was about 80\% -- 90\%.  This means that,  if we have  two sets of VLT observations in the same band 
with approximately equal conditions  and exposure lengths and detect  
a faint object with $S/N\sim 3$ in one of them,  the probability of losing this
object in backgrounds  at the second set  of observations by pure count statistic   
is about 10\% - 40\%.   Accounting for not quite homogeneous conditions for
the 2001 and 2003 periods,  a conservative estimate for losing the suggested
pulsar optical counterpart  at the expected
position  at the 2001 epoch is about   50\%.

  In addition, a  slightly higher background, e.g., 
due to the closeness of a nearby bright object east of the pulsar path (cf. Fig.~2)
or  to a  small increase in the interstellar extinction (e.g., by $\Delta E(B-V)\sim   
0.04$)  for the line of sight towards   the 2001  pulsar
position  can drown the candidate in backgrounds  in the  2001 B image.

The pulsar optical brightness variation 
 cannot be
  excluded  also, since it is known to show  
a pulse nulling  phenomenon in the radio range (e.g., Bhat et
al.~\cite{bhat07} and references therein).  { It spends   up to 15\% of its time in the null
state,  and shows epochs 
when  its radio flux
changes 
 by a factor of  two. There are still no direct
simultaneous  radio and optical observations  of pulsars.  However,  one
cannot rule out a link between pulsar activities in these domains since both
emissions are governed by  complicated NS magnetosperic processes.
The 2001 observing period of one week duration is much shorter than
the 2003 one spanning 8 months and a chance to meet a comparable flux depression
phase that might completely hide the counterpart
in the 2001 epoch is higher.

 We also cannot  exclude the
possibility that  $B$ is simply a time variable background 
object or a part of an extended unresolved variable
feature that includes  $A$ and $B$ as its relatively brighter regions. 
 Another possibility is that $A$ and $B$ represent the same object 
displaced between 2001 and 2004 epochs by its own proper motion.
However, our 
analysis has shown (Sect.~4.1) that  $A$ and $B$ are
likely to be two single objects and $A$ does not show any significant
variability or  proper motion towards the position of $B$
to explain the variability of the latter.  
We have also found no other  variable objects in the pulsar neighborhood at 
the time base of three years.  Possible variability of $B$ 
 can be checked
only by further observations.  If they will show that  $B$  survives and moves 
consistently with the pulsar this will be a strong proof that we see the real optical couterpart.

 A  good positional coincidence of $B$ with the
candidate X-ray pulsar counterpart  suggests that we see
the same object in the optical and X-rays.             
Its color index, $B-R \lesssim0.5$,
is  compatible with  the indices of Vela pulsar,  PSR B0656+14, Geminga, and
PSR B0950+08,
 which all have  $B-R$ in a narrow range of  0.46 - 0.6   
 (Mignani \&  Caraveo  \cite{Mignani2001}; Shibanov et al. \cite{Shib06};  Zharikov et al. \cite{zhar2004}).

 Let us assume that the  optical object $B$ and
 the respective X-ray source are  indeed the  pulsar 
 \object{\psr}. The spectral fit  of the Chandra/ACIS-S  
 data  of the X-ray counterpart candidate 
 by an absorbed power law (Kargaltsev et al.~\cite{Karga06}) 
  yields  the non-thermal  X-ray luminosity of the pulsar $L_X=5.01(+4.30~/-2.41)\times 10^{28}$ in  
  the 2-10 keV range at the distance of 357 pc, or $\log L_X = 28.70(28)$. 
 The corresponding X-ray efficiency is $\log \eta_X
 =-3.24(28)$, which  is not exceptional and lies near the efficiency 
 range of  the two  other  old  pulsars, \object{PSR B1929+10} and \object{PSR B0950+08},  detected in the
 optical  and X-rays  
(Zharikov et al.~\cite{zhar2004, zhar2006}).

\begin{figure}[t]
\setlength{\unitlength}{1mm}
\resizebox{12cm}{!}{
\begin{picture}(100,100)(0,0)
\put (0,0)   {\includegraphics[width=70mm, clip]{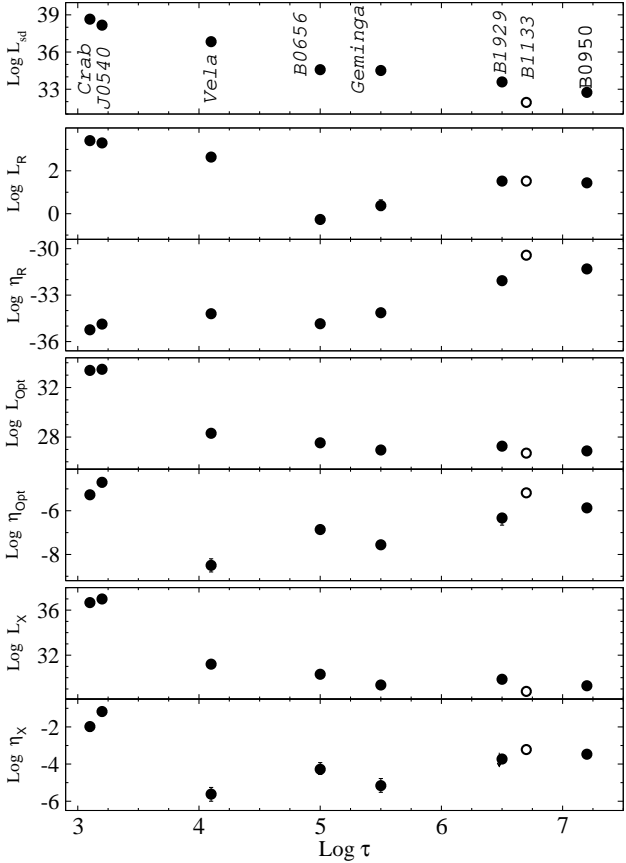}}
\end{picture}
}
\caption{ The evolution of the pulsar radio, optical and X-ray luminosities, 
and respective efficiencies with the characteristic age
$\tau$. The top panel shows corresponding spin-down luminosities.
 % (Zharikov et al. \cite{zhar2005}).  
  }
\label{f:fig4} 
\end{figure}
%%%%%%%%%%%%%%%%%%%%%%%%%%%% end Fig 4 %%%%%%%%%%%%%%%%
%%%%%%%%%%%%%%%%%%%%%%%%%%%%% Fig 5 %%%%%%%%%%%%%
\begin{figure}[t]
\setlength{\unitlength}{1mm}
\resizebox{6.5cm}{!}{
\begin{picture}(95,75)(0,0)
\put (0,0)   {\includegraphics[width=125mm, clip]{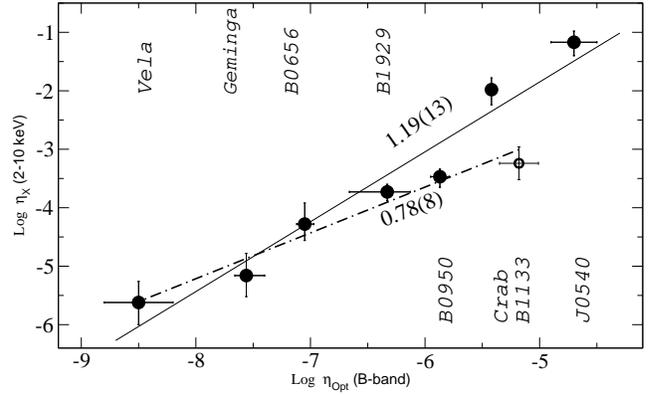}}
\end{picture}
}
\caption{Relations between the optical
$\eta_{Opt}$ and X-ray $\eta_X$ efficiencies in 
the B-band  and  2-10 keV ranges.
 Solid line shows the best linear regression fit  for 
a sample of  pulsars from Zharikov et al.~(\cite{zhar2006}). 
The dot-dashed line is the similar fit only for the set of the pulsars 
that  includes  \object{PSR B1133+16}  counterpart candidate and excludes
 the youngest  \object{Crab} and \object{PSR J0540-69}. Numbers 
 are  the line slopes  with their errors in
 brackets.}
\label{f:fig5} 
\end{figure}
%%%%%%%%%%%%%%%%%%%%%%%%%%%%%%%%%%%%%%%%%%%%%%%%%% end Fig 5 %%%   
 Using $\eta_X$ and the empirical relation between the non-thermal optical and X-ray efficiencies   
 of ordinary pulsars reported by Zharikov et al.~(\cite{zhar2004, zhar2006}),
 we would expect to find  the \object{\psr} optical counterpart with  a magnitude in a range 
 of  about 29--30 and  with the optical efficiency $\log\ \eta_{Opt} \approx -6.2$. The proposed 
 candidate  is a magnitude brighter and the respective 
 optical luminosity and efficiency,  $\log L_B= 26.76(17)$ and $\log\ \eta_{Opt} =-5.18(17)$,  
 are also about one order of magnitude higher than 
 is expected from the reported  
 relation.  However,  this  may simply reflect  
 inaccuracies  in  the  empirical relation, which is  based on a very limited
 sample of the optically detected pulsars.

To study how  this  discrepancy can modify  the relation and the evolution tendencies   
considered by  Zharikov et al.~(\cite{zhar2004,
zhar2006}), 
we included  
the proposed candidate in the sample. The results are shown in Figs.~\ref{f:fig4} and  \ref{f:fig5}. 
As seen (Fig.\ref{f:fig4}), our candidate  does not invalidate  
the previous conclusion regarding the high optical and X-ray efficiencies of old pulsars, 
which become comparable to those of the young Crab-like pulsars. 
At the same time, its position on the $\eta_{Opt}$--$\eta_X$ plane significantly deviates 
from the previous relation  (Fig.~\ref{f:fig5}).  Two of its  neighbors, 
\object{PSR B1929+10} and \object{PSR B0950+08}, also show apparent deviations. 
However, if we exclude 
from the sample 
the two youngest pulsars, the \object{Crab pulsar} and \object{PSR B0540-69}, we obtain a new relation 
that  fits the rest pulsars nicely, including the  \object{PSR B1133+16} candidate.  
This is not a surprise, since  noticed Zharikov et al.~(\cite{zhar2004})   that
 the young pulsars are distinct from the older ones by their  
X-ray and optical efficiencies and  displacement on the $\eta_{Opt}$--$\dot{E}$ and 
$\eta_X$--$\dot{E}$ planes. Therefore, in both respects the proposed candidate is not outstanding. 
That can be considered as  additional evidence  that it is the real counterpart of \object{\psr}.

Despite the high transverse velocity of \object{\psr}, 
we did not detect any developed Balmer bow shock nebula of the pulsar. 
 This suggests   low ISM density  in the pulsar
neighborhood  that is compatible with  
its high galactic latitude  $\approx$69$^o$,
low  $E(B-V)$=0.04 towards the pulsar,  and  a small hydrogen column
density $N_H$=(1-2)$\times 10^{20}$ cm$^{-2}$ (Kargaltsev et
al.~\cite{Karga06}) estimated from the measured  pulsar dispersion measure
of 4.86 cm$^{-3}$ pc  and the spectral fit of the candidate X-ray
counterpart. 
Nevertheless, the emission  detected  at 
about the $3\sigma$ detection limit in the pulsar position error ellipse 
in the H$_{\alpha}$ image
(object $C$)  may come out from the brightest part
of the bow shock  head  { and/or from a clump of the ISM with a higher density.  
The low S/N ratio does not allow us to conclude this  confidently.  
The absence of object $C$ in
our  $B$ and  $R$ broad band  images prabably  excludes its
background 
origin.  Accounting for deep $B$ and $R$ magnitude upper
limits and   a high galactic latitude with low $E(B-V)$,
object $C$   certainly cannot 
be  a star from our Galaxy.  Any background extragalactic object  is also very unlikely since 
it would  be  detected  first of all  in  our  deep broad band images.  

 Instead of the bow shock, 
we   note two marginal but likely spatially correlated extended features in the H$_{\alpha}$ emission and X-rays,  
which are reminiscent of  a ``tail''  behind the pulsar.
 If  the X-ray and the H$_\alpha$  tails  are indeed 
associated with the pulsar, this would present the first example of when the pulsar tail is detected simultaneously 
in the H$_\alpha$ and X-rays.  The combinations of the X-ray tails and H$_\alpha$ bow shocks for a few pulsars 
 have been reported in the literature (see, e.g.,~Zavlin
 \cite{Zav2006} for a short review). 
  Most
recently such a tail has been reliably detected in X-rays 
with the XMM behind  \object{PSR B1929+10} (Becker et al.~\cite{Becker}), 
which is also a member of the  sample considered above.
However, despite these efforts,  no 
counterparts of the tails have been detected so far in the
optical range to our knowledge.
  The formation of the H$_\alpha$ tails 
 does not look unreasonable when a compressed pulsar wind
 behind a supersonically moving NS  cools via X-ray 
 emission that   ionizes the ambient matter on larger spatial
 scales, which then recombines producing H$_\alpha$ photons. An apparent offset of the X-ray tail 
 candidate from the H$_\alpha$ elongated structure in our case 
 (cf. the images with contours in the H$_\alpha$-R and X-ray panels in Fig.\ref{f:fig3})    
 is consistent with such a scenario.

The suggested  candidate 
optical and X-ray counterparts of \object{\psr} and its tail can be  easily verified by 
a followup imaging of the field in the optical and X-rays on the basis of a few years. 
Given the pulsar proper motion, the candidates, if they are  the real counterparts, 
will be shifted from the detected positions roughly to the north by
about 1\farcs4--1\farcs8 by the end of 2008. 
Such a shift  can be reliably measured using  the subarcsecond spatial resolutions 
of 8m-class optical telescopes
and the Chandra X-ray observatory.

\acknowledgements  

 We are grateful  to the anonymous referee for many useful
comments and suggestion that  allowed us  to improve
this paper considerably.  
This work  has been partially supported   by CONACYT  48493  and PAPIIT IN101506
 projects,   RFBR   (grants 05-02-16245 and 05-02-22003),  FASI (grant NSh-9879.2006.2), 
 and Fondecyt 1070705.   
 We used the USNOFS Image and Catalogue Archive
operated by the United States Naval Observatory, Flagstaff Station
(http://www.nofs.navy.mil/data/fchpix/).  
This work is based  on observations made with the European Southern Observatory
 telescopes obtained from the ESO/ST-ECF Science Archive Facility.
The Munich Image Data Analysis System is developed and maintained by the European Southern Observatory.


\begin{thebibliography}{99}
\bibitem[2006]{Becker} Becker, W., Kramer, M., Jessner, A.  2006,  ApJ, 245, 1421  
\bibitem[2007]{bhat07} Bhat, N.D., Gupta, Y., Kramer, M., et al.~2007,  A\&A, 462, 257
\bibitem[2002]{Brisken} Brisken, W.F., Benson, J.M., Goss, W.M., et al.  2002, \apj, 571, 906 
\bibitem[2006] {GaeSlan}Gaensler, B. M., Slane, P. O. 2006, Annual Review of Astronomy \& Astrophysics, 44, 17 
\bibitem[2000]{Gallant} Gallant Y.,  et al. 2000, ESO/VLT program 66.D-0069(A)
\bibitem[1993]{cordes}Cordes, J. M., Romani, R. W., Lundgren, S. C. 1993, Nature, 362, 133
\bibitem[2004]{kar2004} Kargaltsev, O., Pavlov, G. G., Romani, R. 2004, \apj, 602, 327
\bibitem[2006]{Karga06}  Kargaltsev, O., Pavlov, G. G., Garmire, G. P. 2006 \apj, 636, 406 
\bibitem[1992]{Landolt}Landolt, A.  1992, \aj, 104, 340
\bibitem[2001]{Mignani2001} Mignani, R., Caraveo, P. 2001, A\&A, 376, 213
\bibitem[2002]{Mignani2002}Mignani, R., De Luca, A., Caraveo, P., Becker, W. 2002, \apj, 580, L143 
\bibitem[2003]{Mignani03}Mignani, R., Manchester, R. N., Pavlov, G. G. 2003, \apj, 582, 978
\bibitem[2005]{Mignani2005}Mignani, R. 2005,  Proceeding of the NATO/ASI Conference: The Electromagnetic Spectrum of Neutron Stars. Held in Maramaris (Turkey), June 7-18, 2004, astro-ph/0502160
\bibitem[1991]{newberry} Newberry, M. V. 1991, \pasp, 103, 122
\bibitem[1995]{Fukugita}Fukugita, M., Shimasaku, K., Ichikawa, T. 1995, \pasp, 107, 945
\bibitem[1996]{Pavlov}Pavlov, G. G., Stringfellow, G. S., Cordova, F. A. 1996, \apj, 467, 370
\bibitem[1993]{Percival}Percival, J. W.,  Bigg, J. D., Dolan, J. F., et al. 1993, \apj, 407, 276
\bibitem[1998]{Schlegel} Schlegel,  D.,  Finkbeiner,  D.,  Davis, M. 1998, \apj, 500, 525
\bibitem[2006]{Shib06} Shibanov, Yu.,  Zharikov,  S., Komarova, V.  2006 A\&A, 448, 313
\bibitem[2004]{Zav2004}Zavlin, V. E., Pavlov, G. G. 2004, ApJ, 616, 452
\bibitem[2006]{Zav2006} Zavlin, V. E. 2006,  Proceedings of  "Isolated Neutron Stars: from the Interior to the Surface" (April 24-28, 2006) 
- eds. D. Page, R. Turolla \& S. Zane,  to appear in ASpS (arXiv:astro-ph/0608210)  
\bibitem[2002]{zhar2002} Zharikov, S., Shibanov, Yu.,  Koptsevich, A.  et al. 2002,  A\&A, 394, 633
\bibitem[2004]{zhar2004} Zharikov, S.,  Shibanov, Yu., Mennickent R.,  et al. 2004, A\&A, 417, 1017
\bibitem[2006]{zhar2006} Zharikov, S., Shibanov, Yu., Komarova, V. 2006, AdSpR, 37, 1979 
\end{thebibliography}
\end{document}